\documentclass[fleqn,usenatbib]{mnras}

\usepackage{newtxtext,newtxmath}
\usepackage[T1]{fontenc}

\usepackage{graphicx}	% Including figure files
\usepackage{amsmath}	% Advanced maths commands
\usepackage{amssymb}	% Extra maths symbols
\usepackage{xspace}	
\usepackage{tablefootnote}

\newcommand{\unit}[1]{\ensuremath{\mathrm{\,#1}}\xspace}
\newcommand{\kms}{\unit{km\,s^{-1}}}
\newcommand{\masyr}{\unit{mas\,yr^{-1}}}
\newcommand{\gaia}{{\it Gaia}\xspace}
\newcommand{\dhel}{\ensuremath{D_{\mathrm{hel}}}}
\newcommand{\teff}{\ensuremath{T_{\mathrm{eff}}}}
\newcommand{\feh}{\unit{[Fe/H]}}
\newcommand{\code}[1]{\textsc{#1}\xspace}
\DeclareMathOperator*{\argmax}{arg\,max}

\newcommand{\SSSSS}{${S}^5$\xspace}

\title[S5-HVS1]{
The Great Escape: Discovery of a nearby 1700 km/s star ejected from the Milky Way by Sgr A*
}

\author[Koposov et al.]{
\parbox{\textwidth}{
\Large
Sergey~E.~Koposov,$^{1,2}$\thanks{Email:skoposov@cmu.edu}
Douglas~Boubert,$^{3}$
Ting~S.~Li,$^{4,5,6,7,8}$
Denis~Erkal,$^{9}$
Gary~S.~Da~Costa,$^{10}$
Daniel~B.~Zucker,$^{11,12}$
Alexander~P.~Ji,$^{4,6}$
Kyler~Kuehn,$^{13,14}$
Geraint~F.~Lewis,$^{15}$
Dougal~Mackey,$^{10,16}$
Jeffrey~D.~Simpson,$^{17}$
Nora~Shipp,$^{18,8,7}$
Zhen~Wan,$^{15}$
Vasily~Belokurov,$^{2}$
Joss~Bland-Hawthorn,$^{15,16}$
Sarah~L.~Martell,$^{17,16}$
Thomas~Nordlander,$^{10,16}$
Andrew~B.~Pace,$^{19}$
Gayandhi~M.~De~Silva,$^{14,16}$
and Mei-Yu~Wang$^{1}$
\begin{center} (\SSSSS collaboration) \end{center}
}
\vspace{0.4cm}
\\
\parbox{\textwidth}{
$^{1}$ McWilliams Center for Cosmology, Carnegie Mellon University, 5000 Forbes Ave, Pittsburgh, PA 15213, USA\\
$^{2}$ Institute of Astronomy, University of Cambridge, Madingley Road, Cambridge CB3 0HA, UK\\
$^{3}$ Magdalen College, University of Oxford, High Street, Oxford OX1 4AU, UK\\
$^{4}$ Observatories of the Carnegie Institution for Science, 813 Santa Barbara St., Pasadena, CA 91101, USA\\
$^{5}$ Department of Astrophysical Sciences, Princeton University, Princeton, NJ 08544, USA\\
$^{6}$ Hubble Fellow\\
$^{7}$ Fermi National Accelerator Laboratory, P.O.\ Box 500, Batavia, IL 60510, USA\\
$^{8}$ Kavli Institute for Cosmological Physics, University of Chicago, Chicago, IL 60637, USA\\
$^{9}$ Department of Physics, University of Surrey, Guildford GU2 7XH, UK\\
$^{10}$ Research School of Astronomy and Astrophysics, Australian National University, Canberra, ACT 2611, Australia\\
$^{11}$ Department of Physics \& Astronomy, Macquarie University, Sydney, NSW 2109, Australia\\
$^{12}$ Macquarie University Research Centre for Astronomy, Astrophysics \& Astrophotonics, Sydney, NSW 2109, Australia\\
$^{13}$ Lowell Observatory, 1400 W Mars Hill Rd, Flagstaff,  AZ 86001, USA\\
$^{14}$ Australian Astronomical Optics, Faculty of Science and Engineering, Macquarie University, Macquarie Park, NSW 2113, Australia\\
$^{15}$ Sydney Institute for Astronomy, School of Physics, A28, The University of Sydney, NSW 2006, Australia\\
$^{16}$ Centre of Excellence for All-Sky Astrophysics in Three Dimensions (ASTRO 3D), Australia\\
$^{17}$ School of Physics, UNSW, Sydney, NSW 2052, Australia\\
$^{18}$ Department of Astronomy \& Astrophysics, University of Chicago, 5640 S Ellis Avenue, Chicago, IL 60637, USA\\
$^{19}$ George P. and Cynthia Woods Mitchell Institute for Fundamental Physics and Astronomy, and Department of Physics and Astronomy, Texas A\&M University, College Station, TX 77843, USA\\
}
}

\date{Accepted XXX. Received YYY; in original form ZZZ}

\pubyear{2019}

\begin{document}
\label{firstpage}
\pagerange{\pageref{firstpage}--\pageref{lastpage}}
\maketitle
 
\begin{abstract}
We present the serendipitous discovery of the fastest Main Sequence hyper-velocity star (HVS) by the Southern Stellar Stream Spectroscopic Survey ($S^5$).  The star S5-HVS1 is a $\sim 2.35$\,M$_\odot$ A-type star located at a distance of $\sim 9$\,kpc from the Sun and has a heliocentric radial velocity of $1017\pm 2.7$\,\kms without any signature of velocity variability. The current 3-D velocity of the star in the Galactic frame is $1755\pm50$\,\kms. When integrated backwards in time, the orbit of the star points unambiguously to the Galactic Centre, implying that S5-HVS1 was kicked away from Sgr A* with a velocity of $\sim 1800$\,\kms and travelled for $4.8$\,Myr to its current location. This is so far the only HVS confidently associated with the Galactic Centre.  S5-HVS1 is also the first hyper-velocity star to provide constraints on the geometry and kinematics of the Galaxy, such as the Solar motion $V_{y,\odot}= 246.1\pm 5.3$\, \kms or position $R_0=8.12\pm 0.23$\,kpc. 
The ejection trajectory and transit time of S5-HVS1 coincide with the orbital plane and age of the annular disk of young stars at the Galactic centre, and thus may be linked to its formation. With the S5-HVS1 ejection velocity being almost twice the velocity of other hyper-velocity stars previously associated with the Galactic Centre, we question whether they have been  generated by the same mechanism or whether the ejection velocity distribution has been constant over time.
\end{abstract}


%%%%%%%%%%%%%%%%%%%%%%%%%%%%%%%%%%%%%%%%%%%%%%%%%%

%%%%%%%%%%%%%%%%% BODY OF PAPER %%%%%%%%%%%%%%%%%%

\section{Introduction}

%In a relaxed stellar system, the velocity distribution of stars should be of the %order of the virial velocity of the system. In the absence of collisions,

Throughout the last 100 years of studying our Galaxy, there was always a prominent niche in identifying fast moving stars on the sky or in 3-D. One of the first studies of high-velocity stars was the PhD thesis by \citet{Oort1926} who put a boundary between high velocity and low velocity stars at 63\,\kms.
Initially, the searches for fast moving stars were focused on using the proper motions \citep{vanMaanen1917,Lyuten1979} because these were easier to obtain in larger numbers than radial velocities. Due to the fact that the tangential velocities are distance dependent, these searches provided us with some of the first large samples of nearby and Milky Way (MW) halo stars  \citep{Barnard1916,Eggen1967,Eggen1983}. 

When larger numbers of radial velocities began to be analysed in the 1950s--1960s \citep{Kennedy1963} the term "high velocity star" was used to refer to the stars with space velocities of $\gtrsim 100$\kms \citep{Keenan1953}, where those stars were mostly MW stellar halo stars \citep{Eggen1962}. Around the same time, another type of high velocity object emerged -- the runaway OB stars \citep{Blaauw1954}. These stars did not have extreme space velocities, but instead were just  offset from the expected velocity of the disk by 100-200\kms. Some stars were later found in the MW halo \citep{Greenstein1974} with velocities up to 200\,\kms. The mechanism proposed for the formation of such high velocity stars involves either a supernovae explosion in a binary \citep{Blaauw1961} or ejection due to encounters in clusters \citep{Poveda1967}. 

For a while these pathways seemed to be the most promising for creating fast moving stars in the Galaxy with velocities potentially up to the escape speed.  However \citet{Hills1988} proposed an entirely new mechanism of creating fast moving stars with velocities of $1000$\,\kms and above (labelled hyper-velocity stars, HVS) by interaction of a stellar binary with a super-massive Black Hole (SMBH) in the centres of galaxies. This mechanism was almost forgotten until the early 2000s,  when \citet{Yu2003} analysed the ejection mechanism from single and binary SMBHs and \citet{Brown2005} identified a star in the Milky Way halo at a distance of $40-70$\,kpc with a total velocity of $\sim$ 700 \kms, well above the escape velocity at such a distance.  This discovery spurred a renewed interest in hyper-velocity stars \citep{Edelmann2005,Hirsch2005,Heber2008,Przybilla2008} and led to dedicated searches, resulting in multiple new HVS \citep{Zheng2014,Huang2017,Irrgang2019} and candidate HVS \citep[see][for a detailed overview and more references]{Brown2015}.

The most recent part of the story is the arrival of \gaia\ data 
\citep{GaiaMission2016}, in particular Data Release 2 \citep{GaiaDR2} that provided high accuracy proper motions, and thus enabled new discoveries \citep{Shen2018}, potential discoveries \citep{Marchetti2018,Hattori2018b,Bromley2018,Boubert2019} as well as detailed studies of the HVS origins  \citep{Boubert2018,Brown2018,Irrgang2018,Erkal2019}. One of the key conclusions from these studies is that despite the large number of HVS candidates, only a handful of these appear to be actually unbound from the Galaxy and consistent with ejection from the Galactic Centre (GC).

Whilst the extreme speed of several of the HVS in the outer halo is seemingly unexplainable without the Hills mechanism, the uncertainties on their distances and proper motions are such that they cannot be tracked back precisely to the GC. The most convincing association to date is the star J01020100-7122208, identified by \citet{Massey2018} as a bound runaway star that in a particular choice of potential tracked back to the Galactic Centre; however, the low 3-D velocity of $296 \kms$ does not preclude a more standard origin. There is not yet an example of an HVS that unequivocally tracks back to the GC, and thus no smoking gun for a GC Hills mechanism ejection. The power of HVS as probes of the Galactic potential \citep{Gnedin2005} and the orbit of the Sun \citep{Hattori2018a} is contingent on an unambiguous GC origin, and thus it is of paramount importance that such a smoking gun is found.

In this paper we present the discovery of a new nearby unbound HVS that can be unambiguously traced back to the Galactic Centre. The star is named S5-HVS1 as it was found in the Southern Stellar Stream Spectroscopic Survey (\SSSSS, \citealp{Li2019}).

The structure of the paper is as follows.
In Section~\ref{sec:data} we briefly introduce the \SSSSS survey data that was used to identify the S5-HVS1 star and the search for HVS stars in \SSSSS data. In Section~\ref{sec:properties} we look at the spectroscopic and photometric properties of S5-HVS1. In Section~\ref{sec:kinematics} we analyse the kinematics of the star and its possible origin in the Galaxy. In Section~\ref{sec:gc_origin}  we focus on the Galactic Centre as a source of S5-HVS1 as well as inferences we can make on the Galactic potential, distance and velocity of the Sun with respect to the Galactic Centre. We discuss S5-HVS1 in more detail in  Section~\ref{sec:discussion} by comparing it to other HVS, as well as examining HVS ejection mechanisms. Our conclusions are given in Section~\ref{sec:conclusions}.

\section{Data} 
\label{sec:data}

The \SSSSS project is a survey devoted to the observation of stellar streams in the Southern Hemisphere \citep{Li2019}. The survey is being conducted on the 3.9\,m Anglo-Australian Telescope (AAT) with the Two-degree Field (2dF) fibre positioner feeding the AAOmega dual arm spectrograph \citep{Lewis2002,Sharp2006}. \SSSSS uses low (580V, $R\sim1300$) and high (1700D, $R\sim 10000$) resolution gratings in the blue and red wavelength ranges respectively, covering the Balmer break region ($3800<\lambda<5800$\AA) in the blue and IR Calcium triplet ($8400<\lambda<8800$\AA) in the red.
The survey is ongoing, but by early 2019 it had observed 110 fields spread across $\sim$330 square degrees and $\sim$40000 targets. For details we refer the reader to the \citet{Li2019} paper, while providing here only the key aspects of the survey. 

\SSSSS is primarily targeting stellar stream candidate members, selected based on photometric information from the Dark Energy Survey (DES) DR1 \citep{Abbot2018} and proper motion and parallax information from \gaia\ DR2 \citep{GaiaMission2016,GaiaDR2}. To fill all the 392 fibres of the spectrograph other target classes are observed, including low-redshift galaxy candidates, white dwarfs (WDs), and metal-poor stars, etc. The survey specifically targets blue stars that could be either Blue Horizontal Branch (BHB) stars, Blue Stragglers (BS) or RR Lyrae stars at a large range of distances. The selection used by \SSSSS for the BHB/BS stars is $-0.4<(g-r)<0.1$ and ${\tt parallax}<3*{\tt parallax\_error}+0.2$, combined with the star-galaxy separation criteria using $\tt astrometric\_excess\_noise$ quantities from \gaia \citep{Lindegren2018A,Koposov2017} and $\tt wavg\_spread\_model$ quantities from DES \citep[see Eq. 1-3 in][]{Li2019}. At the time of writing the \SSSSS catalogue contains spectra of $\sim 3500$ blue faint objects. While many of them end up being quasars \citep[see][]{Li2019},  $\gtrsim 2200$ of them are likely BHB/BS/WD stars.

The data processing of the \SSSSS data includes standard data reduction steps by the AAT pipeline, followed by spectral modelling by the \code{rvspecfit}\footnote{\url{http://github.com/segasai/rvspecfit}} software in order to determine the radial velocities and stellar atmospheric parameters. 

\subsection{HVS star search}

While identifying hyper-velocity stars was not a main goal of the \SSSSS survey, the catalogue of radial velocities (RVs) and spectral fits was inspected for stars with velocities larger than $800$\,\kms. The majority of objects with such high RVs were spurious measurements caused by either sky subtraction residuals and/or low signal-to-noise spectra, however the search identified a single bright ($G\sim 16$) star with the \gaia\ DR2 {\tt source\_id} 6513109241989477504 and $(\alpha,\delta)=(343.715345\degr, -51.195607\degr)$, located in the field of the Jhelum stellar stream, a new stellar stream found in the DES~\citep{Shipp2018}. This star had a confident radial velocity measurement of $\sim 1020$\,\kms, making it one of the fastest moving stars known in the Galaxy. The radial velocity of this star alone, irrespective of the distance, is enough to make the star unbound to the Galaxy \citep[see e.g.][]{Kafle2014}.
We label this star S5-HVS1\footnote{S5-HVS1 was previously photometrically identified as a candidate field BHB star by \citet{Christlieb2005} and given the designation HE~2251--5127.}. In the next sections we focus on the detailed measurements of S5-HVS1 properties: spectroscopic, photometric and kinematic. 

\section{S5-HVS1 properties}
\label{sec:properties}

In this section we discuss the key spectroscopic properties of S5-HVS1 as determined from AAT data, as well as all available photometric data.
The summary of these measurements is presented in Table~\ref{tab:hvspar}. 

\begin{table}
    \caption{The measured parameters of the hyper-velocity star S5-HVS1. The top part of the Table refers to the measurements from previous surveys, while the bottom one summarises the measurements presented in the paper. HRV is the heliocentric radial velocity. $\dhel$, $D_{\rm hel,GC}$ are heliocentric distance constraints without and with the Galactocentric origin assumption respectively. $V_{\rm GSR}$, $V_{\rm GSR,GC}$ are the inferred Galactic standard of rest (GSR) velocities of S5-HVS1 determined without and with the Galactocentric origin assumption respectively. $V_{\rm ej,GC}$ is the expected ejection speed from the Galactic Centre. $\mu_{\alpha,\rm pred} \cos \delta$, $\mu_{\delta,\rm pred}$ are the predicted proper motions of S5-HVS1 based on the Galactocentric origin.
}
    \centering
    \begin{tabular}{c|c|c}
    \hline
        Parameter & Value & unit \\
        \hline
        \gaia RA& $343.715345$ & deg  \\
        \gaia Dec &  $-51.195607$& deg \\ 
        \gaia\ DR2 {\tt source\_id} & 6513109241989477504 & \\
        \gaia\ $\mu_\alpha \cos \delta$ & $35.328 \pm 0.084$ & \masyr \\
        \gaia\ $\mu_\delta$ & $0.587 \pm 0.125$ & \masyr \\
        \gaia Parallax & $-0.042 \pm 0.091$& mas \\
        \gaia RUWE & 1.06284 & \\
        ${\it E(B-V)}_{\rm SFD}$ & $0.00721$ & \\
        \gaia G   & $16.0211$ & mag \\
        DES $g,r,i,z$& $15.90, 16.16, 16.40,16.53$ & mag \\ 
        $G_{\it BP}-G_{\it RP}$ & $-0.0082 \pm 0.0066$ & mag\\
\hline
        HRV & $1017.0 \pm 2.7$ & \kms \\
        \teff & $9630\pm110$ &  K \\
        $\log g$ & $4.23\pm0.03$ &  dex \\
        \feh & $0.29\pm0.08$ &  dex \\
        $\log_{10} D_{\mathrm{hel}}/1\,{\rm kpc}$ &  
        $0.936 \pm  0.015$ &  \\  
        $V_{\rm GSR}$ &  $1755_{-45}^{+55}$& \kms \\
        $V_{\rm GSR,GC}$  &  $1717.4\pm 3.5$& \kms  \\
        $V_{\rm ej,GC}$ &  $1798.6\pm 3.1$& \kms\\
        $D_{\rm hel,GC}$ & $ 8884\pm11$ & pc\\
        $\mu_{\alpha,{\rm pred}} \cos \delta$ & $35.333\pm 0.080$ &  \masyr \\
        $\mu_{\delta,{\rm pred}} $ & $0.617\pm 0.011$ & \masyr \\
        \hline
    \end{tabular}
    \label{tab:hvspar}
\end{table}

\subsection{Spectroscopy}
\label{sec:spectra_analysis}

\begin{figure*}
	\includegraphics{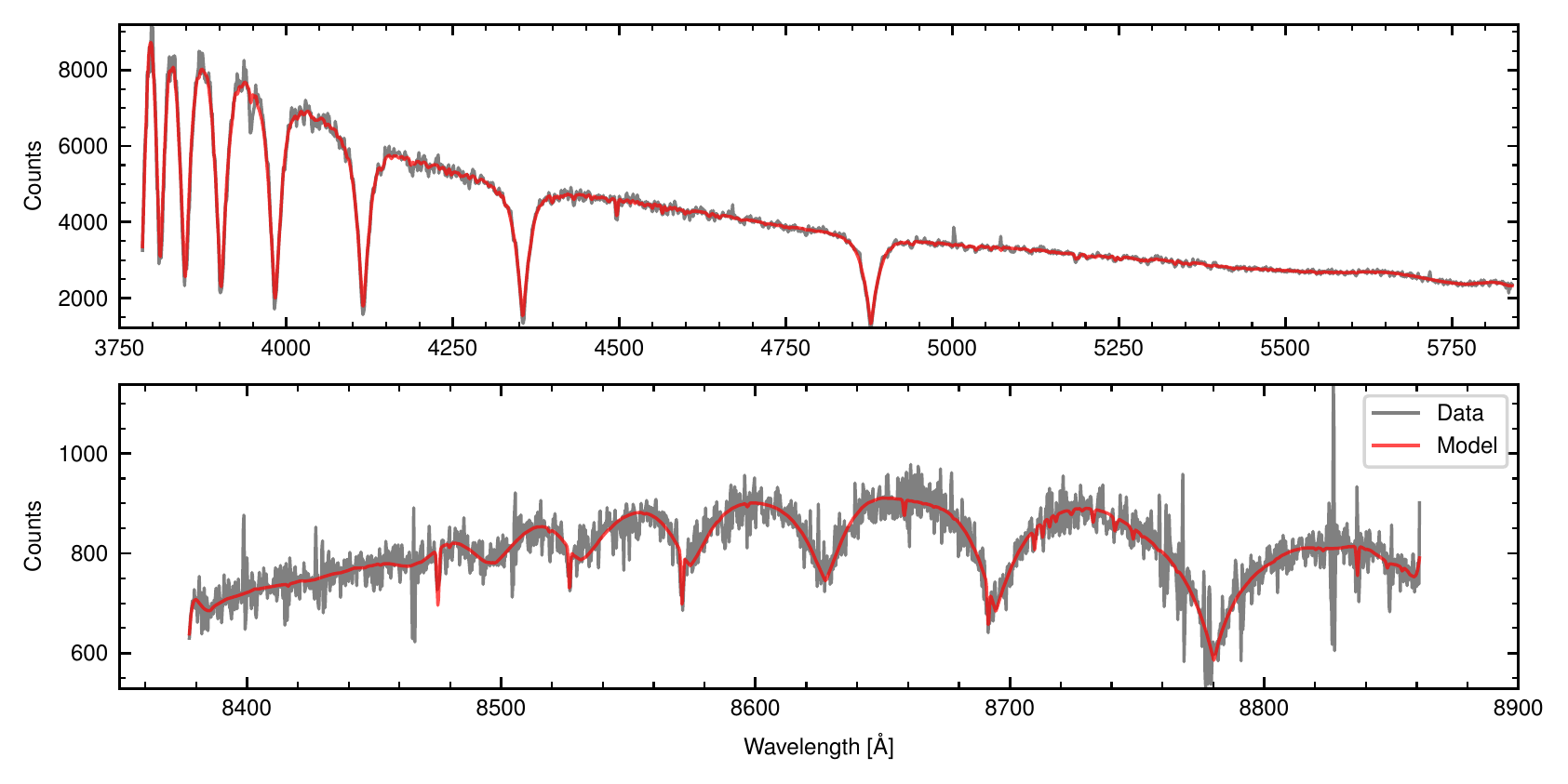}
    \caption{The blue and red spectra of S5-HVS1. The grey lines show the spectra from \SSSSS AAT observations, obtained using 580V (top panel) and 1700D (bottom panel) AAT gratings.  The red lines show the best fit model based on interpolated spectral templates from the PHOENIX library \citep{Husser2013}, which was determined by simultaneous fitting to the blue and red data.}
    \label{fig:specfit}
\end{figure*}
The star S5-HVS1 was observed for the first time at the AAT as part of regular \SSSSS observations of the Jhelum stellar stream with the 580V and 1700D gratings on 2018 August 1. The total exposure time was 2 hours split into three individual exposures. The combined, reduced spectra for S5-HVS1 are shown in Figure~\ref{fig:specfit}. Based on the spectra, the star appears to be a hot A-type star with prominent broad Balmer and Paschen series and several metal lines like Ca II H/K and Mg II (4481\AA) in the blue and Calcium triplet in the red. 

Although the stellar spectra of S5-HVS1 in both the blue and red arms were analysed as part of the regular \SSSSS processing \citep[see][]{Li2019}, the analysis treated the blue and red arms separately.  
For this paper, however, we analyse the blue and red parts of spectra simultaneously in order to better constrain stellar atmospheric parameters.
The fitting of stellar spectra is analogous to the procedure described in the \SSSSS overview paper and uses the \code{rvspecfit}, but instead of considering the likelihood function of the red arm or blue arm data separately, we combine them.
Specifically the model for the stellar spectrum uses a combination of global Radial Basis Function interpolation and 
local linear N-d interpolation of spectra from the PHOENIX-2.0 library~\citep{Husser2013} together with a multiplicative polynomial to deal with the fact that the observed spectra were not flux calibrated \citep[see][]{Koposov2011}.

\begin{multline*}
\operatorname{Model}(\lambda|\log g, \teff, \feh, V) =  \left( \sum\limits_{i=0}^{n_\mathrm{p}} 
 a_i \lambda^i \right)\ \times \\ 
 {\mathcal T}\left(\lambda\left[1+\frac{V}{c}\right], \log g, \teff, \feh\right)
\end{multline*}
Here $\lambda$ is the wavelength, the $ {\mathcal T}(\lambda, \log g, \teff, \feh)$ is the interpolated stellar template, V is the radial velocity, $a_i$ are fitted coefficients and $n_{\mathrm{p}}$ is the degree of the multiplicative polynomial used to correct for continuum normalisation\footnote{Since the blue arm part of the spectra has a much larger wavelength calibration uncertainty \citep[see][]{Li2019}, when we fit for stellar atmospheric parameters we allowed for a small RV offset between blue and red arms.}. The parameters of the model for the star were then sampled using the parallel tempering Ensemble sampling algorithm \citep{Goodman2010,Foreman_Mackey2013} to determine uncertainties.
We adopted non-informative uniform priors on all parameters \citep[i.e. contrary to][we did not use the \teff\ prior based on the colour of the star]{Li2019}.

The red curve in Figure~\ref{fig:specfit} shows the  best-fit spectral model corresponding to the maximum likelihood set of parameters. The stellar atmospheric parameters are effective temperature $\teff  = 9630\pm110$\,K, surface gravity $\log g=4.23\pm 0.02$, and high stellar metallicity $\feh=0.29\pm 0.08$. We note though that the posterior is bi-modal with two modes at $(\teff,\feh)\sim (9500$K$,0.25)$ and $ (9700$K$,0.4)$. This is likely caused by the limitations of the adopted stellar atmosphere grid and interpolation procedure, as the resolution of the PHOENIX grid is $0.5$\,dex in $\log g$ and $\feh$ and $\sim$ $200-500$\,K in $\teff$. Because of this, the uncertainties on the stellar atmospheric parameters should be mostly systematic. Despite that, the measured surface gravity of S5-HVS1 strongly suggests that the star is a Main sequence A-type star as opposed to a Blue Horizontal Branch star with $\log g\lesssim3.5$\footnote{We remark that formally the star lies on the BHB side of the $\log g$, $\teff$ distribution shown on Figure 11 of \citet{Li2019}. However the analysis presented in \citet{Li2019} relied only on 1700D data as opposed to combination of 580V and 1700D data that we use here, and is therefore somewhat on different scale.}

While we determined the atmospheric parameters for S5-HVS1 from simultaneous fitting of the red and blue spectra separately from main \SSSSS data processing, the radial velocity measurement for the S5-HVS1  that we will use comes from the main \SSSSS catalogue. The RVs in the catalogue rely only on the red arm of the spectra, as its wavelength calibration and stability are much better controlled due to a higher spectral resolution and the presence of large number of skylines in the science spectra.  As discussed in detail in \citet{Li2019}, the radial velocities and their uncertainties measured  in \SSSSS have been validated with both repeated observations and observations of \gaia RVS and APOGEE stars. The uncertainties on the radial velocities also take into account  the systematic error floor in our observations of $\sim 0.6$\kms. The heliocentric radial velocity measured for S5-HVS1 by \SSSSS is $1017.0\pm 2.7$\kms. The blue arm spectrum provides an independent velocity measurement with a similar value albeit with much larger error-bar $1017\pm23$\kms. 

\subsection{Radial velocity variability}
\label{sec:rv_variab}

The radial velocity of S5-HVS1 is extreme and thus we must consider the possibility that it is due to binary motion. To check this hypothesis, we re-observed the star almost 8 months after the first observation. The first repeated observation was done on 2019 April 6 (MJD $58579.78$; i.e., 240 days after the first observation) again using AAT 2dF spectrograph in the same configuration as in the \SSSSS survey. We ensured that S5-HVS1 was assigned to a different fibre and plate from our 2018 observation to rule out any possible fibre-specific effects. The observations were performed in twilight and had an exposure time of only $2 \times 900$s and therefore were of lower S/N than standard \SSSSS data\footnote{On 2019 April 6, this star was above airmass $\sim2$ for only 10 min before astronomical twilight.}. Consequently the red (1700D) spectrum was not usable, but fortunately the 580V blue spectrum had S/N $\sim 3$ and we were able to measure a velocity of $V = 1017\pm 24 $\kms which is consistent within uncertainties with the original measurement. 

We also carried out a further re-observation of S5-HVS1 on 2019 April 26 (MJD 58599.78) using the WiFeS integral field spectrograph \citep{Dopita2010} on the ANU 2.3m telescope at Siding Spring Observatory.  The instrumental setup employed the B3000 grating that gives resolution $R\sim3000$ and wavelength coverage of 3500--5600\AA.  Two 900s exposures were obtained and the combined reduced spectrum yielded a heliocentric velocity of $1005 \pm 15$ \kms, which is entirely consistent with the other observations.  In addition, model atmosphere spectral fits to the WiFeS flux-calibrated spectrum yielded an effective temperature of approximately $10,000$\,K, and more importantly, a surface gravity $\log g$ of $4.5$, confirming the main sequence star nature of S5-HVS1.

From these  additional observations spread over a few months, we can convincingly rule out a binary origin of the high velocity of S5-HVS1, because high binary orbital velocities $\gtrsim 100$\kms are only expected in binaries with high masses and short periods.  It is still possible that S5-HVS1 is part of a long-period binary with a small orbital velocity that is undetectable in a period of $\sim$ a year, but this orbital motion would be negligible compared to the observed RV. Therefore most of the observed radial velocity must be caused by the motion through the Galaxy.

\subsection{Photometry}
\label{sec:photometry}

S5-HVS1 was targeted by \SSSSS as a blue star with $-0.4<g-r<0.1$, which makes it a possible BHB or BS. In this Section we assess the photometric properties of S5-HVS1 by collecting its photometry across multiple wavelengths and fitting these data with an isochrone model.

\begin{figure}
	\includegraphics[width=\columnwidth]{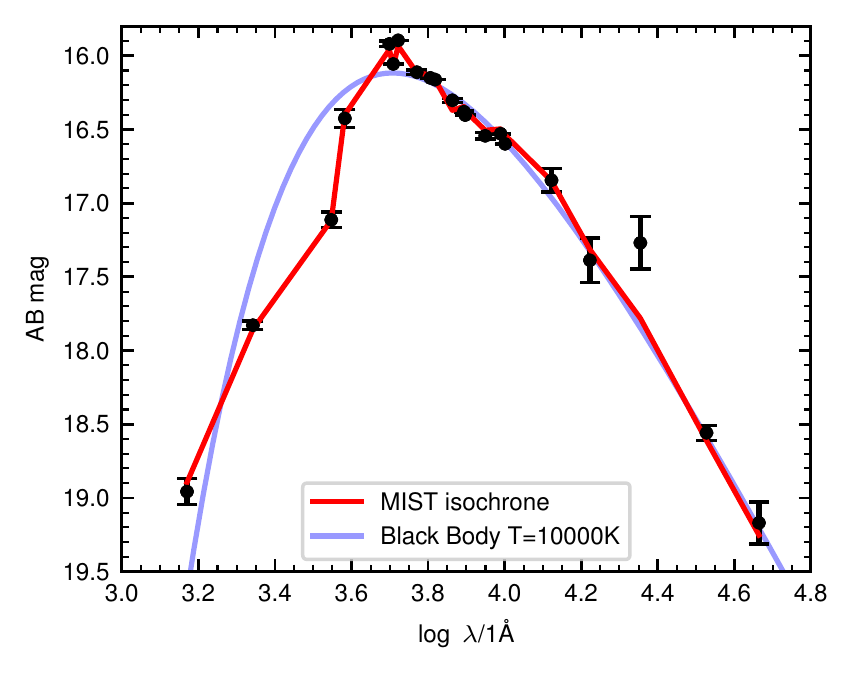}
    \caption{
    Spectral energy distribution (SED) of S5-HVS1 from GALEX, \gaia, SkyMapper, DES,
    2MASS and WISE photometry. The blue curve is the black-body spectrum with temperature of 10000\,K. The red line shows the SED from the best-fit MIST isochrone model. The magnitudes in the data and model were not extinction corrected.
    }
    \label{fig:sed}
\end{figure}

As S5-HVS1 is quite bright, \gaia\ $G\sim 16$, it is detected in a large number of different surveys. Here we take the data from DES DR1 \citep{Abbot2018}, 2MASS \citep{Skrutskie2006}, AllWISE \citep{Wright2010}, SkyMapper DR1.1 \citep{Wolf2018}, GALEX \citep{Martin2005,Bianchi2017} and \gaia DR2 \citep{Brown2018, Evans2018}. Figure~\ref{fig:sed} shows all S5-HVS1 magnitudes (converted when needed from Vega to AB magnitude system) as a function of the effective wavelength of the corresponding filter with standard errors. The SED is clearly indicative of a hot star with temperature $\sim10000$\,K. The red line shows the photometry from the best fit isochrone model in the observed filters that we describe below. The blue line shows a black body spectrum with a temperature of $10000$\,K.

{\renewcommand{\arraystretch}{1.2}% for the vertical padding
% latex high tech for you... 
\begin{table}
    \caption{The parameters measured from fitting MIST isochrones to the S5-HVS1 SED (Model P) and by combining SED constraints with spectroscopic constraints (Model SP).
    }
    \centering
    \begin{tabular}{c|c|c|c}
    \hline
    Parameter & Value & Value & unit \\
     & Photometric & Spectro-Photometric & \\
    \hline
    Mass & $1.90^{+0.25}_{-0.28}$ & $2.35^{+0.06}_{-0.06}$ &  M$_\odot$\\
    $\log_{10}$ age & $8.36^{+0.32}_{-0.46}$ &  $7.72^{+0.25}_{-0.33}$ & dex\\
    $\feh$ & $-0.2^{+0.2}_{-0.3}$ & $0.3^{+0.1}_{-0.1}$  & dex\\
    m-M & $14.21^{+0.37}_{-0.43}$ & $14.68^{+0.07}_{-0.07}$ & mag\\
    $\sigma_{\rm sys}$ & $0.04^{+0.01}_{-0.01}$ & $0.03^{+0.01}_{-0.01}$ & mag\\
    \hline
    \end{tabular}
    \label{tab:sed_fitres}
\end{table}
} % this is the brace closing the renewcommand arraystretch

\begin{figure*}
	\includegraphics{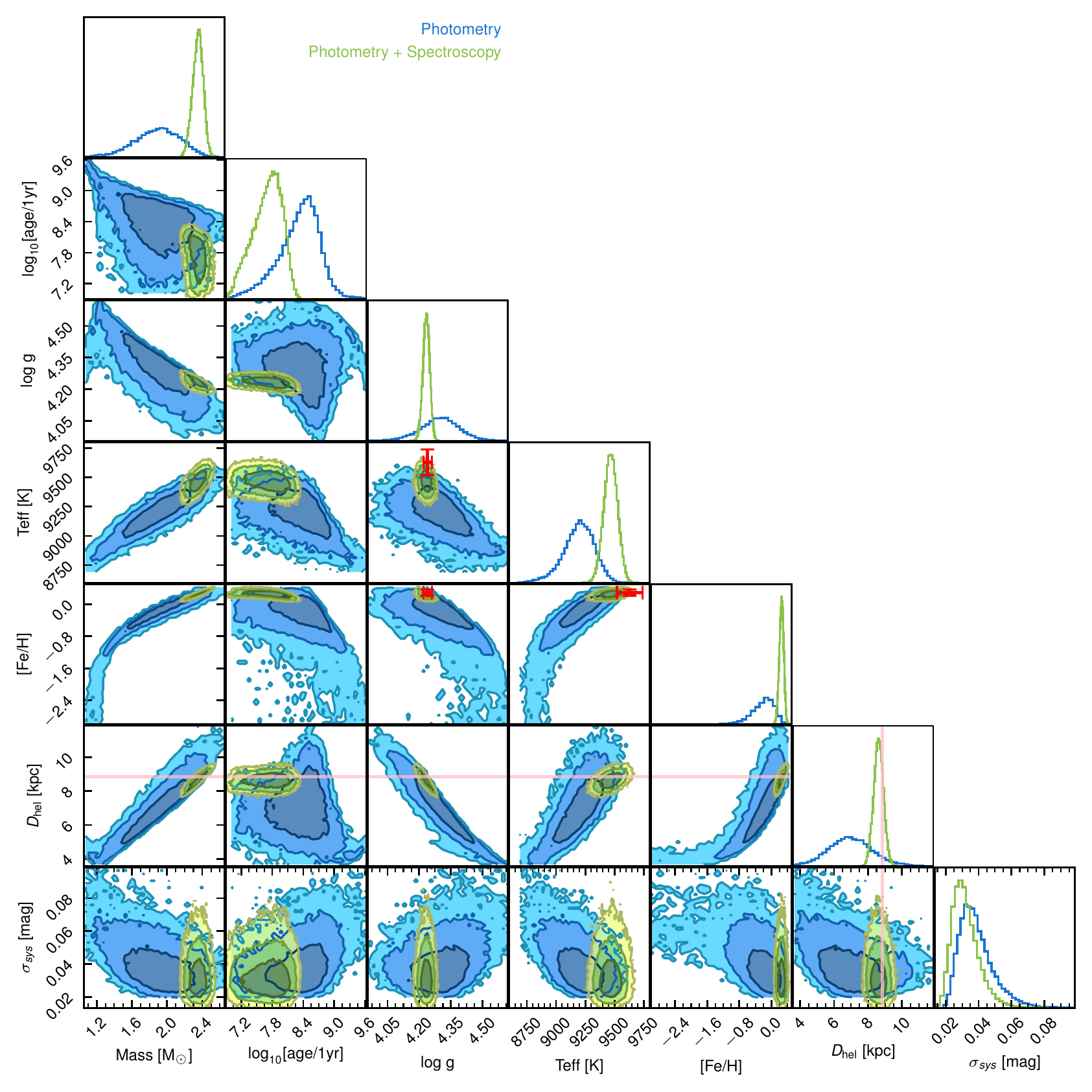}
    \caption{The posterior on stellar parameters of S5-HVS1 from fitting MIST isochrones to the SED data  only (blue) and the SED data combined with the prior on stellar atmospheric parameters from spectroscopic analysis (green). The red points with error-bars are showing the best-fit measurement of stellar atmospheric parameters from the analysis of the AAT spectra using \code{rvspecfit}. 
    The pink lines on several panels identify the  heliocentric distance to the star that is consistent with the Galactocentric origin (see Section~\ref{sec:gc_origin}). The contour levels in the 2-D marginal distributions corresponds to the 68\%, 95\% and 99.7\% of posterior volumes.}
    \label{fig:sedfit_corner}
\end{figure*}
To model the photometry of S5-HVS1 we use the MIST isochrones \citep[][version 1.2]{Dotter2016,Choi2016}  and to interpolate between isochrones we use the \code{isochrones} software \citep[][version 2.0.1]{Morton2015}\footnote{For \gaia $G_{\it BP}$, $G_{\it RP}$ magnitudes we use the band-passes defined by \citet{Weiler2018}.}. The data that we model are the observed magnitudes ${m_{i}}$ where $i$ corresponds to the $i$-th band. The isochrones provide us with absolute magnitudes, surface gravities and effective temperatures as a function of stellar age, mass, metallicity  and band-pass $M(\mathrm{age}, {\mathcal M}, [\mathrm{Fe}/\mathrm{H}], i)$. Assuming Gaussian uncertainties of observed magnitudes, our model is

\begin{equation}
\begin{split}
m_i  & \sim  {\mathcal N}\bigg( M (\mathrm{age},{\mathcal M},[\mathrm{Fe}/\mathrm{H}], i) + \\
 &  + 5 \log_{10} D_{\mathrm{hel}} -5 + k_i E(B-V), \sqrt{\sigma_i^2 + \sigma_{\mathrm{sys}}^2}\bigg),
\end{split}
\label{eq:phot_model}
\end{equation}
where $\sigma_i$ is the uncertainty on the magnitude measurement in band $i$,  $\sigma_{\mathrm{sys}}$ is an additional (systematic) scatter around the model, $D_{\mathrm{hel}}$ is the heliocentric distance to the star, and $k_i$ is the extinction coefficient\footnote{Taken from \url{http://www.mso.anu.edu.au/~brad/filters.html}} in the filter $i$. On top of the purely photometric model described in Eq.~\ref{eq:phot_model} (we label it Model P), we also consider a model (labelled Model SP) where we complement Eq.~\ref{eq:phot_model} with the constraints on $\log g$, $\teff$ and $[\mathrm{Fe}/\mathrm{H}]$ from the spectroscopic analysis (see Section~\ref{sec:spectra_analysis}), assuming they are normally distributed (i.e. we multiply the likelihood by Gaussian  terms for $\log g$, $\teff$ and $[\mathrm{Fe}/\mathrm{H}]$).

We adopt generically uninformative priors for the parameters: uniform distribution on (linear) $\mathrm{age} \sim {\mathcal U}(10^5, 1.2\times10^{10})$, Salpeter IMF prior for the stellar mass from ${\mathcal M}=0.1$\,M$_\odot$ to ${\mathcal M}=5$\,M$_\odot$, uniform prior on metallicity $[\mathrm{Fe}/\mathrm{H}] \sim {\mathcal U}(-4, 0.5)$,
and a uniform prior on distance modulus $5 \log_{10} D_{\mathrm{hel}} -5 \sim {\mathcal U}(10,20)$ corresponding to a $1/D^2$ spatial density prior from $1$\,kpc to $100$\,kpc. For the extinction, we adopt a prior around the \citet{Schlegel1998} value $E(B-V) \sim {\mathcal U}( 0.3 E_{\mathrm{SFD}}, 3 E_{\mathrm{SFD}})$.
The posterior of the model is sampled using the nested sampling \code{MultiNest} algorithm \citep{Feroz2008,Buchner2014}.

The posterior of the model parameters is shown in Figure~\ref{fig:sedfit_corner}; blue contours and curves for Model P and green for Model SP. Focusing on the Model P first, we notice that as expected from photometric only data there are considerable degeneracies between mass, age, metallicity and distance of the star. The summary of parameters for Model P is provided in Table~\ref{tab:sed_fitres}. The age of the star is consistent with a broad range of ages up to $500$\,Myr. The mass of the star is inferred to be $1.9\pm 0.25$\,M$_\odot$. The distance to the star is constrained to be $\log_{10} \frac{D_{\mathrm{hel}}}{1 {\rm kpc}} = 0.836\pm0.083$, putting it in the range of between $\sim 4.5$ and $10$\,kpc from the Sun. We notice that this distance corresponds to a parallax of $\pi_
\mathrm{phot}\sim0.14$\,mas which is consistent within 2 sigma with the negative \gaia\ parallax measurement $\pi_{\gaia}=-0.042\pm 0.091$\,mas that was not used in the fit.  The systematic error for the photometry is determined by the model to be $\sigma_{\rm sys}=0.04\pm0.01$ showing that there is no large discrepancy between isochrone models and data. 

The match between the data and the isochrone model across the wavelengths is well demonstrated by Figure~\ref{fig:sed}.
Red points with error-bars shown on multiple panels of  Figure~\ref{fig:sedfit_corner} mark the parameter values measured from spectroscopic analysis of S5-HVS1 (Section~\ref{sec:spectra_analysis}). The measurements from photometric data only are broadly consistent with the spectroscopic analysis, as the error-bars overlap with the high probability parts of the posterior. Although there is possibly a small discrepancy in temperature of $\sim 200$\,K and/or $\feh$ of $\sim 0.2$\,dex between purely spectroscopic and photometric measurements, we believe this level of disagreement is well within the systematic errors of our spectroscopic and isochrone modelling. 

Since the photometric and spectroscopic analyses are consistent, we also show in the Figure the posterior from the combination of the spectroscopic and photometric analyses (Model SP) as green contours.  As expected, the combination of the datasets shrinks the posteriors considerably, i.e. the combined mass estimate is $2.35\pm 0.06$\,M$_\odot$  and distance estimate is $\log \frac{\dhel}{1 {\rm kpc}}=0.936\pm0.015$. The posterior estimates for these and other parameters from the Model SP are also provided in Table~\ref{tab:sed_fitres}. Throughout the paper we use both the photometric and photometric+spectroscopic sets of estimates, where we will interpret the photometric-only constraints as being more conservative\footnote{In the final stages of preparation of the manuscript, we identified that S5-HVS1 has a distance estimate of $\log_{10} 
\frac{\dhel}{1\,{\rm kpc}}=0.807\pm0.148$ from the \code{StarHorse} code \citep{Anders2019}, which is in very good agreement with our photometric-only measurement.}. As we will discuss in the next section, the kinematics of S5-HVS1 are consistent with ejection from the Galactic Centre if the star has a very specific heliocentric distance of $\sim 8.8$\,kpc. Pink lines on Figure~\ref{fig:sedfit_corner} show the heliocentric distance to the star that is consistent with ejection from the GC (see Section~\ref{sec:gc_origin}), and we remark that this distance agrees perfectly with both the photometric and spectro-photometric analyses.

\begin{figure*}
	\includegraphics{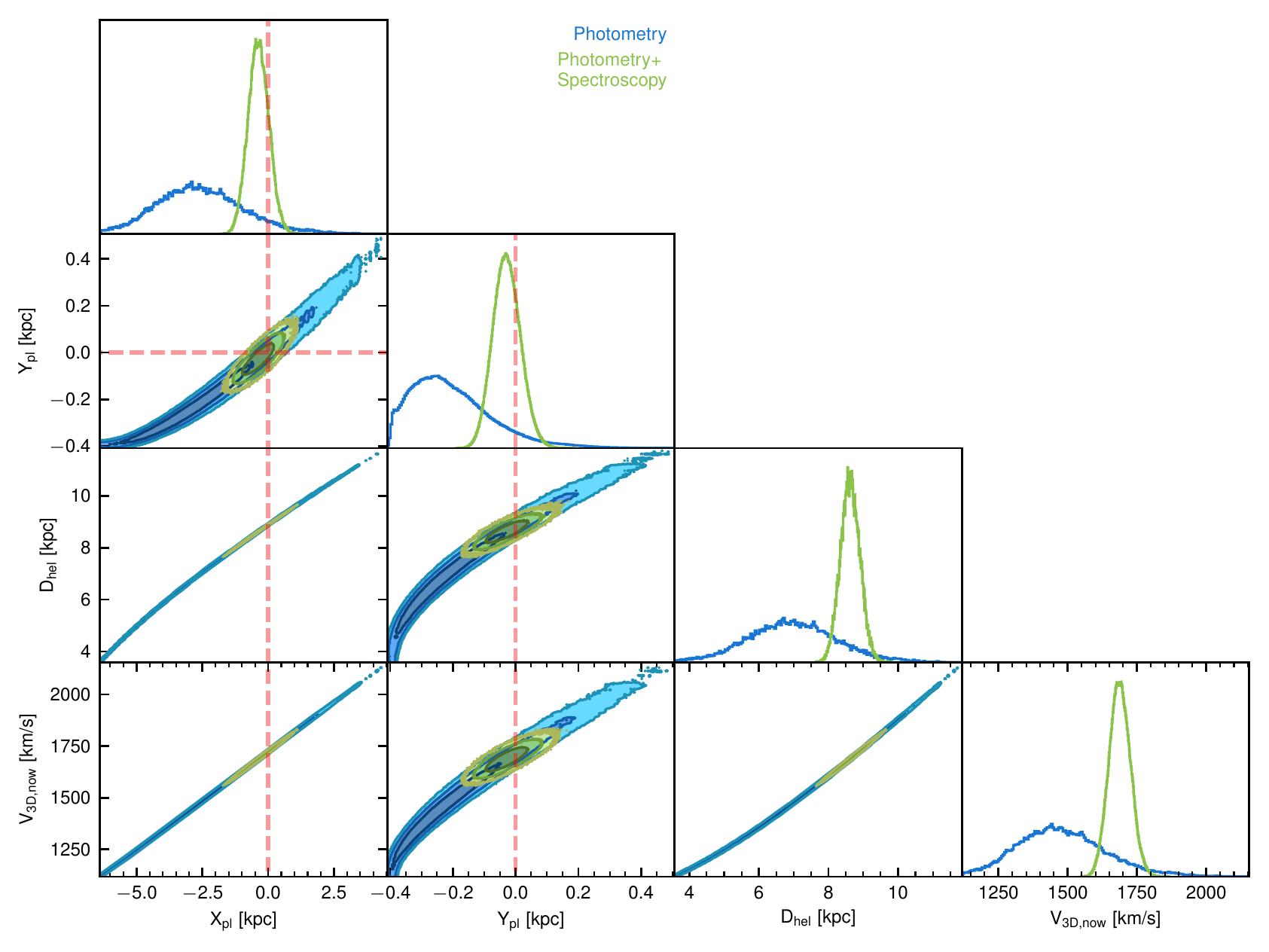}
    \caption{The constraints on possible orbital properties and origin of S5-HVS1, assuming that it was ejected from a point in the Galactic plane. $X_{\mathrm{pl}}$ and $Y_{\mathrm{pl}}$ are the Galactocentric coordinates of the ejection point in the plane. $D_{\mathrm{hel}}$ is the current distance to the star and $V_{\mathrm{3D,now}}$ is the current velocity of the star in the Galactocentric frame. The blue and green contours and curves refer to the posterior that we obtained while using the photometric only distance (Model P) and spectro-photometric distance (Model SP) respectively. We note that the $D_{\mathrm{hel}}$ distributions are exactly the same as in Figure~\ref{fig:sedfit_corner} as we reuse the samples from the SED modelling posterior. The dashed lines identify the location of the Galactic Centre. The lines in the contour plots show the 68\%, 95\% and 99.7\% posterior volumes.}
    \label{fig:orbit_back}
\end{figure*}
While the isochrone modelling performed so far did include the horizontal branch phase, the posterior on the stellar parameters indicates that the photometry of S5-HVS1 is inconsistent with it being a Blue Horizontal Branch (BHB) star. However, it is still worth specifically addressing the possibility that S5-HVS1 is a BHB, because this is quite feasible given the star's colour of $g-r\sim-0.27$ (most BHB stars have colours of $-0.3\lesssim g-r\lesssim 0$, \citealp{Yanny2000}). 
We therefore perform an independent check to assess the BHB hypothesis by looking at measurements of $g-r$ and $i-z$ colours. This colour combination is known to be sensitive to the surface gravity of stars due to the Paschen break contribution to the $z$-band, and therefore allows us to separate BHB from BS/MS stars \citep[see e.g.,][]{Vickers2012,Belokurov2016}. With colours of $(g-r)=-0.27$ and $(i-z)=-0.13$,  S5-HVS1 sits significantly below the line separating the BHB from BS/MS \citep[see right panel of figure 11 and eq. 5 of][]{Li2019} further confirming that S5-HVS1 is a Main Sequence star. Additionally, when looking at the distribution of surface gravities and effective temperatures \citep[left panel of figure 11 of][]{Li2019} S5-HVS1 lies on the BS side of the distribution. Thus for future analysis, unless specified otherwise, we will adopt the Main Sequence based distance constraints determined in this Section. 

\section{Kinematics of S5-HVS1}
\label{sec:kinematics}

The extreme radial velocity of S5-HVS1 as measured from the observed spectra makes it one of the fastest stars known in the Galaxy and thus warrants a detailed investigation of its orbit and origin.
\begin{figure*}
	\includegraphics[height=2.9in]{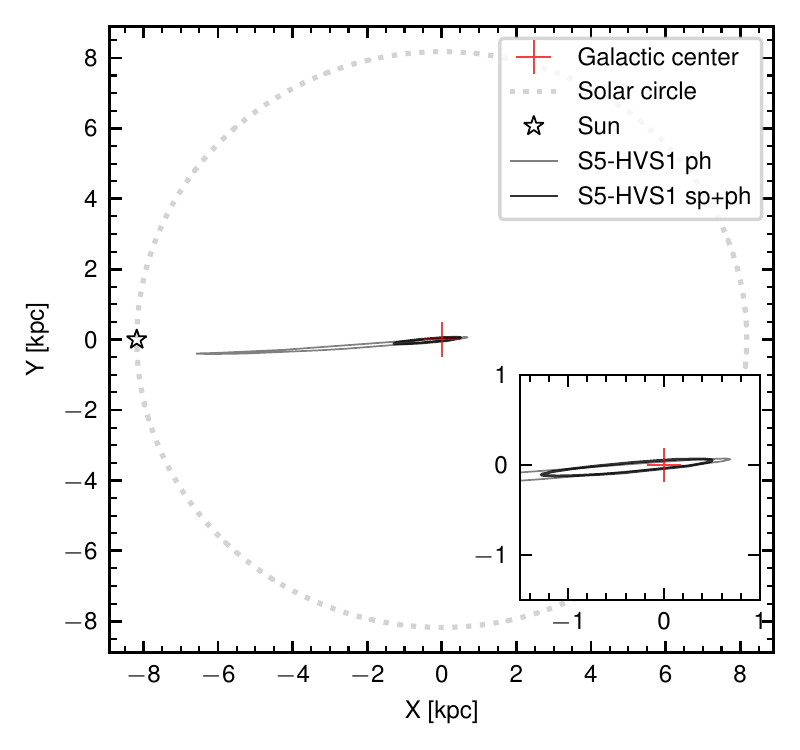}
\hspace{-0.4 cm}
	\includegraphics[height=2.9in]{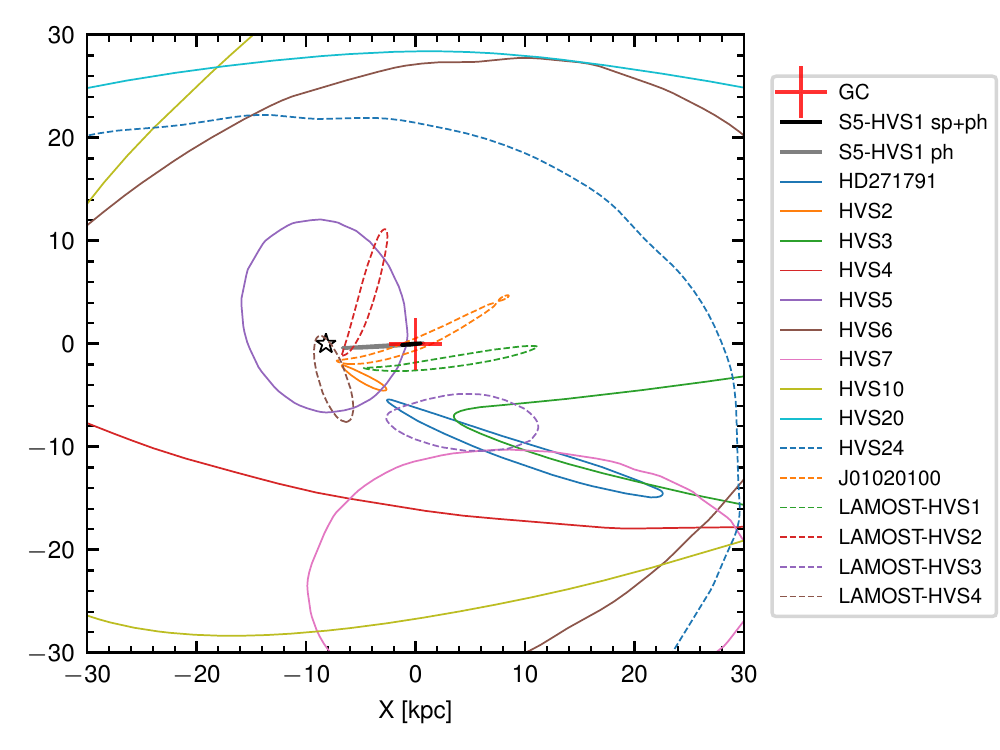}
    \caption{{\it Left panel:} The constraints on the origin of S5-HVS1 in the Galactic plane. The location of the Sun, Solar circle and the Galactic Centre are indicated by a star symbol, grey dotted line, and red cross respectively. The black contour shows the 90\% confidence region of the origin of S5-HVS1 constructed using spectro-photometric distances, while the grey contour shows the constraints if we use less well determined photometric only distances. Both of these contours contain the GC. The small inset shows the central $2.5\times2.5$\,kpc$^2$ region around the GC. {\it Right panel:} 90\% confidence regions in Galactic $X,Y$ for the point of origin of various hyper-velocity stars under the assumption that they come from the Galactic plane. We only included stars with contours that significantly overlap with the $30\times 30$\,kpc$^2$ region shown. The confidence regions for the S5-HVS1 origin are the barely visible grey and black streaks around the GC compared to all other stars.}
    \label{fig:finetuning}
\end{figure*}
Summarizing the phase-space information available for S5-HVS1, the position of the star on the sky is known very precisely, as is the radial velocity. The proper motion of the star is available in the \gaia\ DR2 catalogue and, because of the star's brightness $G\sim 16$, it is also very precise  $(\mu_\alpha \cos \delta ,\mu_\delta) = ( 35.328 \pm 0.084, 0.587 \pm 0.125)$\masyr\footnote{The astrometry of S5-HVS1 does not seem to be affected by any astrometric problems according to the re-normalised unit weight error (RUWE) \citep{Lindegren2018A}, which is $\sim$ 1.06.} .  The only phase space parameter that is poorly constrained is the heliocentric distance, as discussed in Section~\ref{sec:photometry}. This is why we expect that most of the orbital inferences for  S5-HVS1 should show a 1-D degeneracy corresponding to a range of possible heliocentric distances. Even with the more conservative (Model P) distance estimates, it is clear from combining the radial velocity and proper motions that the S5-HVS1 velocity in the Galactic frame is in excess of $\sim 1200$\kms:  $V_{\rm 3D}=1470_{-147}^{+166}$ \kms.

As a first step in modelling the orbit of S5-HVS1, we perform a backward integration of its current phase space coordinates to infer a possible ejection site of the star. Since the current total velocity of S5-HVS1 is at least $ 1200$\,\kms, 
one of the key questions we are interested in is whether the star has been ejected from the Galactic Centre, the MW disk, or some other system such as a globular cluster or satellite galaxy.  While some fast moving stars have been associated with other galaxies like the Large Magellanic Cloud \citep{Edelmann2005,Gualandris2007,Boubert2016,Boubert2017,Irrgang2018,Erkal2019}, in this paper we will focus only on ejection from the MW disk and the Galactic Centre.

To infer a possible ejection point and velocity of S5-HVS1, we integrate the orbit of the star backwards in time in the gravitational potential of the Milky Way until the star intersects the Galactic plane $Z=0$ at the location $X_{\mathrm{pl}},Y_{\mathrm{pl}}$. Throughout the paper when doing orbit integrations, unless specified otherwise, we adopt the gravitational potential from \citet{Mcmillan17}, the distance from the Sun to GC of 8.178\,kpc \citep{Gravity2019}, and Solar velocity of $(U_\odot,V_\odot,W_\odot)=(11.1,245,7.25)\,\kms$ \citep{Schonrich2010,Mcmillan17}. To take into account the observational uncertainties in our inference of the ejection site $X_{\mathrm{pl}},Y_{\mathrm{pl}}$, when integrating back the orbit of S5-HVS1 we sample the observed uncertainties in radial velocity from \SSSSS, proper motion from \gaia\ and the distance posterior derived in Section~\ref{sec:photometry}. The resulting distributions of the Galactic plane ejection coordinates $X_{\mathrm{pl}},Y_{\mathrm{pl}}$ together with current heliocentric distance $D_{\mathrm{hel}}$ and current velocity in the Galactocentric frame $V_{\mathrm{3D,now}}$ are shown in Figure~\ref{fig:orbit_back}. The two sets of distributions shown with green and blue correspond to the photometric only distance (Model P) and spectro-photometric distance (Model SP) constraints. We remark that the current heliocentric distance $D_{\mathrm{hel}}$ distribution is exactly the same as the posterior on $D_{\mathrm{hel}}$ determined in Section~\ref{sec:photometry}. While the Figure shows the Monte-Carlo sampling of uncertainties, it is mathematically equivalent to the posterior distribution of $P(X_{\mathrm{pl}},Y_{\mathrm{pl}}, D_{\mathrm{hel}}, V_{\mathrm{3D,now}}| \mathrm{Data})$ under the model where the star was ejected from MW disk plane (and uninformative priors on $X_{\mathrm{pl}},Y_{\mathrm{pl}}$ and ejection velocity).

As expected, the posterior on the S5-HVS1 ejection point is very elongated (almost one dimensional) due to the negligible uncertainties in all parameters but the heliocentric distance. However, we also see that the usage of spectro-photometric distances alleviates this problem somewhat. The current total velocity of the star in the Galactic rest-frame is constrained to be $V_{\mathrm{3D,now}}= 1470_{-150}^{+170}$\kms for Model P and  $V_{\mathrm{3D,now}}= 1687_{-37}^{+39}$\kms for  Model SP, while the ejection velocity of the star from the Galactic disk is $V_{\mathrm{ej}} = 1550_{-160}^{+190} $\kms for Model P, and $V_{\mathrm{ej}} = 1755_{-44}^{+45}$\kms for Model SP, very similar to the current velocity. The difference between the current velocities and the ejection velocities is small ($\sim 50$\kms) because the impact of the Galactic potential on such a fast moving star is minimal.
The inferred ejection point based on the photometric only distance (Model P) is $X_{\mathrm{pl}}=-2.63_{-1.54}^{+1.72}$\,kpc, $Y_{\mathrm{pl}}=-0.22_{-0.10}^{+0.15}$\,kpc, where the values and uncertainties come from 50\% and 16\%, 84\% percentiles of 1-D marginal distributions. However the constraints on $X_{\mathrm{pl}},Y_{\mathrm{pl}}$ are strongly non-Gaussian and elongated. Most importantly we see that the Galactic Centre $(X,Y)=(0,0)$ (shown on Figure~\ref{fig:orbit_back} by pink dashed lines) is located within the $90$\% probability contour of the $X_{\mathrm{pl}},Y_{\mathrm{pl}}$ distribution. While the peak of the posterior for the ejection point $X_{\mathrm{pl}},Y_{\mathrm{pl}}$ is shifted by 2.5\,kpc from the Galactic Centre, the fact that the very thin probability contour covers the GC is highly informative and suggestive of a GC origin. If we instead consider the contours for Model SP based on spectro-photometric distances we see that the inference of  $X_{\mathrm{pl}},Y_{\mathrm{pl}}$ is significantly tighter   $X_{\mathrm{pl}}=-0.37_{-0.39}^{+0.40}$\,kpc,  $Y_{\mathrm{pl}}=-0.03_{-0.04}^{+0.05}$\,kpc, and thus our backwards integrations point almost unambiguously at the Galactic Centre $(X_{\mathrm{pl}},Y_{\mathrm{pl}})= (0,0)$ as the origin of S5-HVS1\footnote{While the backward orbit integration done here uses the potential of \citet{Mcmillan17}, which does not include the SMBH in the Galactic Centre, we have verified that the constraints on the ejection point $(X_{\mathrm{pl}},Y_{\mathrm{pl}})$ are completely insensitive to the presence or absence of a $\sim$ $4\times 10^6\,$M$_\odot$ BH in the Galactic Centre due to its very small sphere of influence compared to the size of the $(X_{\mathrm{pl}},Y_{\mathrm{pl}})$ contours.}.

To further illustrate the strength of evidence supporting an association of S5-HVS1  with the GC we look at the confidence region of the S5-HVS1 origin and compare it to the Solar circle. The left panel of Figure~\ref{fig:finetuning} shows the 90\% confidence region for S5-HVS1 when relying on spectro-photometric distances (the black contour), and photometric only distances (grey contour). Both of the 90\% confidence limits well encompass the GC. We also see that even the less well constrained Model P contour is extremely thin compared to the Solar circle, suggesting that it would be quite unlikely for it to cover the Galactic Centre by random chance. This conclusion applies even more strongly for the minutely thin contours of Model SP as shown by the black line. To formally quantify the statistical significance of the association of S5-HVS1 with the GC, we can use the posterior on $X_{\mathrm{pl}}$, $Y_{\mathrm{pl}}$ to compute the Bayes factor \citep[see e.g.][]{Trotta2007} between the hypothesis that the star comes from the GC vs. that it comes from a random point in the Galactic disk. To do this we have to adopt a prior on $X_{\mathrm{pl}},Y_{\mathrm{pl}}$ for the Galactic disk origin hypothesis. We use the exponential distribution with a scale length of $2.15$\,kpc, to match the distribution of stellar mass in the disk \citep{bovy13}. With this prior we can then use the Savage-Dickey ratio \citep{verdinelli95} to evaluate the Bayes factor of the two hypotheses: GC and disk. 

$$K=\frac{P(\mathrm{GC}|\mathrm{Data})}{P(\mathrm{disk}|\mathrm{Data})}\frac{\pi(\mathrm{disk})}{\pi(\mathrm{GC})}= \frac{P(X_{\mathrm{pl}},Y_{\mathrm{pl}}=0,0|\mathrm{disk},\mathrm{Data})}{\pi(X_{\mathrm{pl}},Y_{\mathrm{pl}}=0,0|\mathrm{disk})}$$.

The Bayes factor is $K=81$ when we use photometric distances (Model P), and $K=354$ for the spectro-photometric distances (Model SP). This constitutes strong (Model P) or overwhelming (Model SP) evidence in favour of the Galactic Centre origin. In the calculation, we assumed the same (uniform) priors over ejection velocities, direction and travel time in both hypotheses. An intuitive explanation of a Bayes factor of 354 is that if before observing S5-HVS1 the odds ratio in favour of the GC origin vs. disk origin was 50/50, then after observing S5-HVS1 we would have to update the odds to be 354/1.

The evidence that S5-HVS1 is coming from the GC is almost definitive and is much stronger than for any other hyper-velocity star we know. To illustrate this we take the list of stars from \citet[][augmented with LAMOST-HVS4 from \citealp{Li2018}, and J01020100 from \citealp{Massey2018}]{Boubert2018}, and perform the calculation of the ejection point $X_{\mathrm{pl}},Y_{\mathrm{pl}}$ within the plane of the disk (identically to that performed on S5-HVS1), given the existing observational constraints on those stars (position, distances, proper motions and radial velocities). The right panel of Figure~\ref{fig:finetuning} shows the 90\% confidence contours for a subset of the stars where those contours overlapped significantly with the $30\times30$\,kpc$^2$ region (for many stars in the list, e.g. {HVS1}, the contours are larger than the whole plot). The figure shows that there are stars that could be associated with the GC based on their phase space coordinates (for instance, HVS20 and HVS24), but their confidence region of origin includes the whole Milky Way disk as well. Among other stars with tighter constraints on the point of origin, only J01020100 \citep{Massey2018} seems to cover the GC (but note that  J01020100 is believed to be a disk runaway due its meagre velocity of 296 \kms). The contour showing the region of origin of S5-HVS1 around the Galactic Centre is almost invisible compared to the other stars.

Concluding this section, based on strong orbital evidence  that points at a region of $\sim 50 \times 1000$\,pc$^2$ around the GC as the origin of S5-HVS1 (Figure~\ref{fig:finetuning}) and the large velocity of S5-HVS1 $V_{\mathrm{3D,now}} \sim 1500-1700$\,\kms that is impossible for a disk runaway star of $\sim 2$\,M$_\odot$ \citep[see][]{Tauris2015}, we can conclude that S5-HVS1 was ejected from the Galactic Centre. It is the first star with such a confident identification. In the next section we will analyse what inferences can made based on this assumption.

\section{Galactic centre origin}
\label{sec:gc_origin}

\begin{figure}
    \centering
    \includegraphics{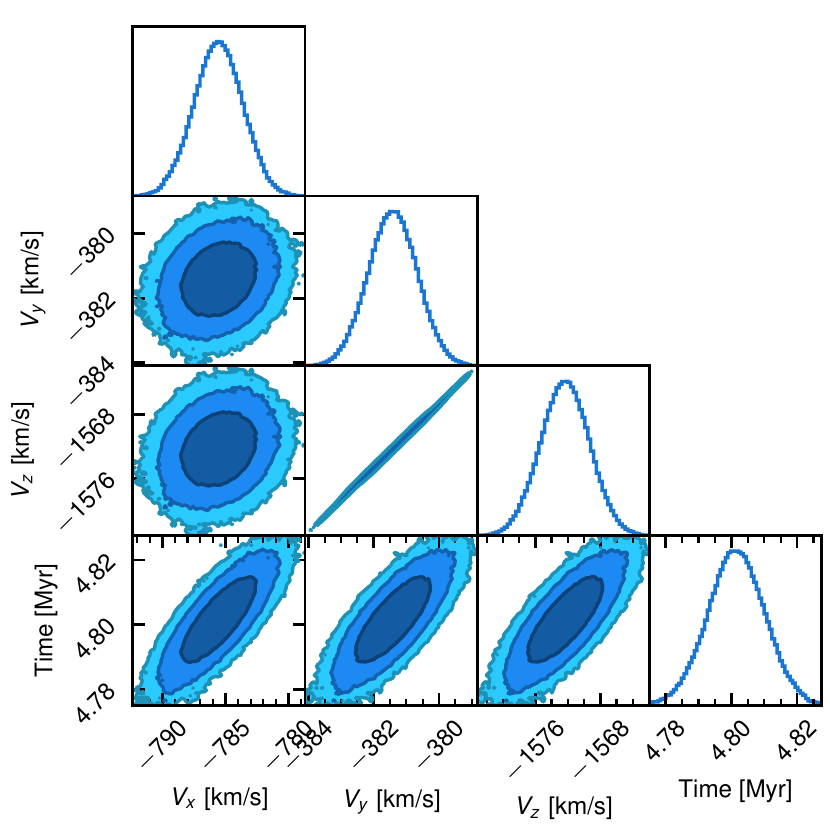}
    \caption{The constraints on the 3 Galactic components of the ejection velocity and travel time to the current location of S5-HVS1, assuming a GC origin. The total ejection velocity is $\sim$ 1800\,\kms oriented downwards from the Galactic plane and somewhat towards the Sun. }
    \label{fig:gc_ejection}
\end{figure}

\begin{figure}
    \centering
    \includegraphics{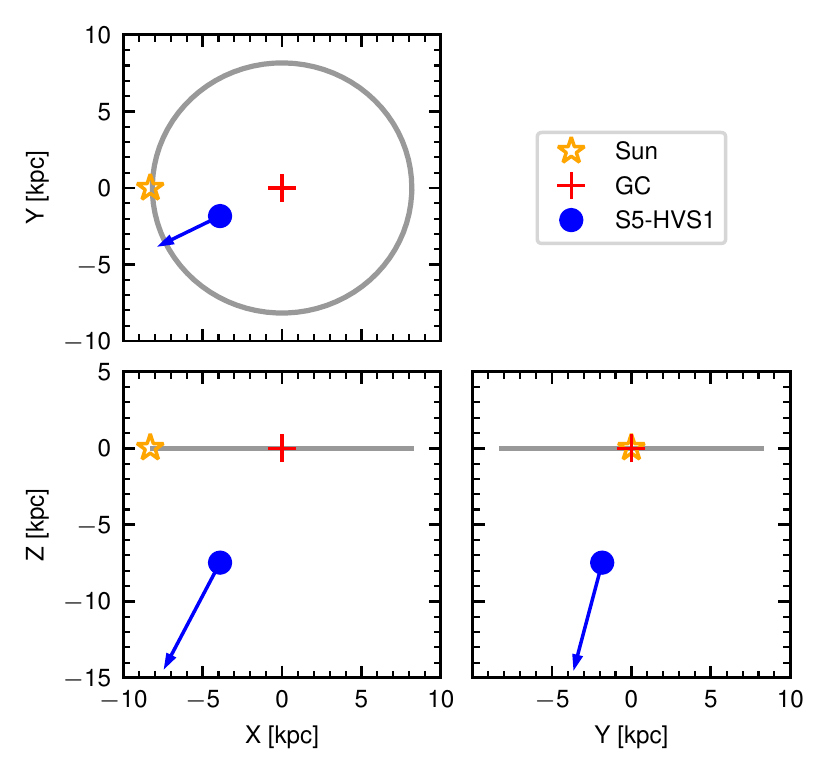}
    \caption{The location and direction of motion of S5-HVS1 in the Galaxy assuming a Galactic Centre origin. Three panels show different projections in Galactic Cartesian coordinates. The location of the Sun, the GC and the Solar circle are marked by the orange star, red cross and grey circle respectively. The arrow shows the direction of S5-HVS1's velocity in the corresponding projection. The length of the arrow corresponds to the distance traveled by the star in 4\,Myr.}
    \label{fig:3d}
\end{figure}

Assuming that S5-HVS1 was ejected from the Galactic Centre, we can now investigate the kinematics of the star further.
First we determine the exact ejection velocity and time of flight from the GC required to match the observations of S5-HVS1, by ejecting the star from the centre of the MW in the potential of \citet{Mcmillan17} (without considering the potential of the SMBH itself). This gives a prediction of $\alpha,\delta,\mu_\alpha,\mu_\delta,RV,D_{\mathrm{hel}}$ as a function of the ejection velocities $V_x,V_y,V_z$ and travel time $T$. We then write a Normal likelihood function using the observed position, distance, proper motion and RV of S5-HVS1 and their uncertainties (we use a Gaussian approximation to the $\log_{10} \dhel$ posterior from Section~\ref{sec:photometry}).
We adopt non-informative uniform priors on all the parameters and then sample the posterior using an ensemble sampler. Figure~\ref{fig:gc_ejection} shows the posterior. This model implies an ejection speed of $1798.6\pm3.1$\kms with the $z$-component of the velocity being the largest and a total travel time from the GC to the current position of $4.801\pm0.009$\, Myr. We note that the constraints on the ejection velocities are now much tighter compared to Figure~\ref{fig:orbit_back}. The reason for this is that postulating that the star is coming from the GC strongly constrains the current distance to S5-HVS1 to be $D_{\mathrm{hel}}=8884\pm 11$\,pc and thus makes our spectro-photometric measurement mostly irrelevant.\footnote{Given a star ejected from the Galactic Centre, it is enough to know accurately just the position of the star on the sky and proper motion to exactly determine its heliocentric distance and radial velocity.}.  We also remark that the measured ejection speed of $1798.6\pm3.1$\kms from the GC was computed while ignoring the potential of the SMBH, and thus represents the ejection velocity outside the sphere of influence of the black hole ($\gtrsim 1$\,pc). The actual ejection speed of the star depends on how close to the BH the ejection happened, and could easily be as high as $\sim$ 8000\,\kms if the ejection happened at a distance of 100\, AU from the BH.
Assuming a GC origin also allows us to improve the proper motion precision from the one delivered by \gaia $\mu_\alpha \cos \delta = 35.333\pm0.081 \masyr$ and  $\mu_\delta=0.617\pm0.011 \masyr$. While the $\mu_\alpha \cos \delta$ precision did not improve much, the error-bar on the predicted $\mu_\delta$ is 8 times smaller than \gaia's.
Since the full phase space position of S5-HVS1 becomes very precise when we adopt the GC origin hypothesis, we can look at the geometric position of S5-HVS1 in the Galaxy. This is shown in Figure~\ref{fig:3d}. We see that as expected, S5-HVS1 is mostly moving downwards away from the disk, and that the Sun, Galactic Centre and S5-HVS1 form an almost equilateral triangle with $\sim 8-9$\,kpc edges.

In this section we will further use the phase space observations of S5-HVS1 to constrain the gravitational potential of the MW, location and kinematics of the Sun in the Galaxy, and assess the possible connection of S5-HVS1 to the stars in the vicinity of Sgr A*.

%\section{Constraints from S5-HVS1}
\subsection{Constraining the position and motion of the Sun}
\label{sec:constraints}

Figure~\ref{fig:finetuning} shows that the association of S5-HVS1 with the GC crucially depends on the relative geometry between the Sun and the Galactic Centre. For example, a small adjustment of the distance from the Sun to the Galactic Centre ($R_0$) could easily shift the high probability contour ${\mathcal P}(X_{\mathrm{pl}},Y_{\mathrm{pl}})$ away from the GC. Therefore assuming that S5-HVS1 originates in the GC constrains $R_0$ and possibly other Galactic parameters.  The idea of constraining the Solar motion as well as the distances to the Galactic Centre have been discussed previously, most notably by \citet{Hattori2018a}. To determine these constraints, we construct a forward model where we eject the star from the GC with velocity $V_x,V_y,V_z$ and let it travel in the Galactic potential of \citet{Mcmillan17} for the time $T$. We then observe it from the Sun located at a distance of $R_0$ from the Galactic Centre and moving with the velocity $U_{\odot}, V_{\odot}, W_{\odot}$\kms (this includes both the speed of the Local Standard of Rest and the peculiar velocity of the Sun). The likelihood of the model is then constructed using the observed 6-D phase measurements of S5-HVS1: position, proper motion, distance and radial velocity. This leads to the following posterior distribution:
 
\begin{eqnarray}
{\mathcal P}(\psi|D)\propto{\mathcal P}(D|V_x,V_y,V_z,T,R_0,U_{\odot},V_{\odot}, W_{\odot})\nonumber\\
\pi(V_x,V_y,V_z)\pi(R_0)\pi(U_{\odot},V_{\odot}, W_{\odot})\pi(T)
\end{eqnarray}
where $\psi$ is the shorthand for all the model parameters $V_x,V_y,V_z,T,R_0,U_{\odot},V_{\odot}, W_{\odot}$.
For this model we focus on constraining $R_0$ and $V_{\odot}$, so we adopt broad uninformative priors on the distance of the Sun to the Galactic Centre $\frac{R_0}{1\, \rm kpc}\, \sim {\mathcal U} (6,9)$, and $\frac{V_{\odot}}{1\,\kms} \sim {\mathcal U} (200, 290)$ and informed Normal priors on the other two components of Solar velocity  $\frac{U_{\odot}}{1\kms} \sim {\mathcal N} (11.1, 0.5)$, $\frac{W_{\odot}}{1\,\kms} \sim {\mathcal N} (7.25, 0.5)$ \citep{Schonrich2010}. For the rest of parameters $V_x,V_y,V_z,T$ we adopt uninformative uniform priors. In principle the model that we have described has a valid posterior that we could sample. However, we have discovered that this posterior is extremely degenerate along one dimension and narrow in another dimension. This is in fact a direct consequence of the elongated contour shape for the constraint on the ejection point $X_{\mathrm{pl}},Y_{\mathrm{pl}}$ seen in Figure~\ref{fig:finetuning}. This contour shape and the fact that simultaneous changes of $V_{\odot}$ and $R_0$ give two degrees of freedom for "moving" the high probability contour in $(X_{\mathrm{pl}}$,$Y_{\mathrm{pl}})$ space while still covering the GC explains the long degeneracy ridge in the posterior. Furthermore the posterior is also extremely narrow along the time axis, as the orbit needs to pass very close to the precisely known observed position on the sky. It turns out that those features of the posterior make it extremely challenging to sample, so we were unable to do it efficiently using either \code{MultiNest}, \code{dynesty}, ensemble or ensemble parallel tempering samplers (\code{emcee}). Our solution to this problem was to adopt an approximation to the posterior where we approximately marginalise over the travel time of the star. 

\begin{eqnarray}
{\mathcal P}(\psi|D) \propto {\mathcal P}(D|V_x,V_y,V_z,T_{\mathrm{max}},R_0,U_{\odot},V_{\odot}, W_{\odot})\nonumber\\
\pi(V_x,V_y,V_z)\pi(R_0) \pi(U_{\odot},V_{\odot}, W_{\odot})
\end{eqnarray}

where $\psi$ is the shorthand for the model parameters $V_x,V_y,V_z,R_0,U_{\odot},V_{\odot}, W_{\odot}$ and
$$ T_{\mathrm{max}} = {\rm \argmax\limits_T}\  {\mathcal P}(D|V_x,V_y,V_z,T,R_0,U_{\odot},V_{\odot}, W_{\odot})$$ 
Thus $T_{\mathrm{max}}$ is the travel time that maximises the likelihood (or approaches the current phase-space constraint of S5-HVS1 the most closely). We find the $T_{\mathrm{max}}$ for each set of parameters by doing 1-D maximisation using the Brent algorithm \citep{Brent2013}. 
The resulting posterior on $V_x,V_y,V_z,R_0,U_{\odot},V_{\odot}, W_{\odot}$ is then sampled using an ensemble sampler with 192 walkers. 

\begin{figure}
\includegraphics{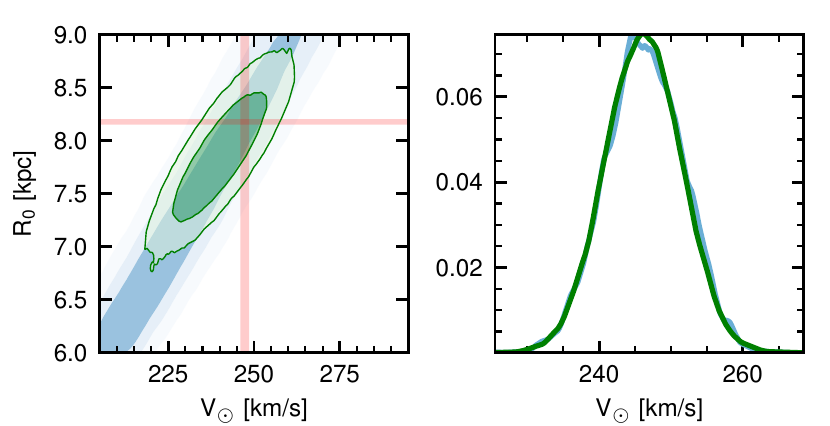}
\caption{{\it Left panel:} The 2-D marginalised posteriors on the heliocentric distance to the GC ($R_0$) and y-component of Solar velocity in the Galaxy as inferred from S5-HVS1. The contours correspond to the 68\% and 95\% posterior volumes. Blue lines on both panels refer to quantities derived from our photometric only distance (Model P), while the green ones refer to the more precise spectro-photometric distances (Model SP). The red bands shows the constraints on $R_0$ and $V_\odot$ from \citet{Gravity2019}. {\it Right panel:} 1-D marginal posteriors on the y-component of Solar velocity in the Galactic rest-frame after adopting a prior on $R_0$ from \citet{Gravity2019}. The inferred value is $V_\odot=246.1\pm 5.3$\kms.}  
\label{fig:geomtry_constraints}
\end{figure}

The left panel of Figure~\ref{fig:geomtry_constraints} shows the 2-D marginalised posterior on two of the parameters -- heliocentric distance to the GC and the Y component of the Solar velocity ($V_\odot$).  As before blue lines correspond to Model P (photometric only distance), and green lines to Model SP (spectro-photometric distance). The red bands show the 1-sigma constraints from \citet{Gravity2019}. As expected the figure shows a degeneracy between parameters which is almost complete when using the less constrained photometric only distances, however with spectro-photometric distances the degeneracy is significantly reduced.  

We note that even with the spectro-photometric distances of S5-HVS1 we cannot strongly constrain both $V_\odot$ and $R_0$ (as the green contours on Figure~\ref{fig:geomtry_constraints} are quite large). However, if we adopt the prior on the Galactocentric distance from \citet{Gravity2019}, we obtain the posterior on $V_\odot$ shown on the right panel of the figure. $V_{\odot}$ is constrained to be $246.1\pm 5.3$ \kms. Those constraints also do not depend significantly on whether  we use spectro-photometric or photometric only distances as we slice the posterior shown on the left panel of the figure across the distance degeneracy. The $V_\odot$ measurement is competitive with and entirely independent from the $247.4\pm1.4$ \kms constraint from \citet{Gravity2019}. If we instead use the prior on $V_\odot$ from \citet{Gravity2019} to constrain the distance to the Galactic Centre (marginalising over the x-axis on Figure~\ref{fig:geomtry_constraints}), we obtain  $R_0=8.12\pm0.23$\,kpc.  

While Figure~\ref{fig:geomtry_constraints} may look somewhat underwhelming compared to the \citet{Gravity2019} measurements, we highlight that our measurement was done with one single star. The shape of the degeneracy in $U_\odot,V_\odot,W_\odot,R_0$ space is specific to the position of the star on the sky, and so if we had a second star then the combined constraints would be significantly more precise and likely comparable in precision to \citet{Gravity2019}\footnote{
In fact in this paper we did not consider determining $U_\odot, W_\odot$, because they are significantly less constrained than $V_\odot$. That can be easily seen because of the shape of the contour of ${\mathcal P}(X_\mathrm{pl},Y_\mathrm{pl})$ on Figure~\ref{fig:finetuning}. The contour is the thinnest in  the direction of solar rotation and is larger by a factor of ten in the $U$ direction.} . Another reason for optimism is that future \gaia data releases and high-resolution spectroscopic follow-up will narrow the uncertainties on the proper motion and distance of S5-HVS1 and thus tighten our constraints on the Solar motion and position in the Galaxy.

\subsection{Constraints on the Galactic potential}

As we showed in the previous section, S5-HVS1 strongly constrains the Galactic position and velocity of the Sun. On top of that we expect that --~under the assumption that the star comes from the GC~-- the observed properties of the star should also constrain the gravitational potential of the MW. This idea was first proposed by \citet{Gnedin2005}, who suggested using individual HVS with accurate phase space positions to measure the MW Dark Matter (DM) halo flattening. This idea has been extended to the modelling of the HVS population as a whole as it is being deflected by the disk and flattened MW halo away from an initial distribution over angles \citep{Yu2007,Contigiani2019}. 

It turns out, however, that while the proximity of S5-HVS1 was essential for precise measurements of its properties and  its detection, the short flight time of $\sim 5$\,Myr from the GC makes the orbit of the star barely sensitive to the MW potential. To gain an intuition for this, it is useful to look at the inferred position where the star crosses the MW plane ${\mathcal P}(X_{\mathrm{pl}},Y_{\mathrm{pl}})$ in Figures~\ref{fig:orbit_back} and \ref{fig:finetuning}, when we were backtracking the current phase space coordinates to the Galactic plane. The reason why the potential could be constrained by the S5-HVS1 is because when we change the gravitational potential, the distribution of $X_{\mathrm{pl}},Y_{\mathrm{pl}}$ changes, and  high probability contours are shifted away from $(X_{\mathrm{pl}},Y_{\mathrm{pl}})=(0,0)$  making such potentials less likely under the hypothesis that the GC is the origin of the S5-HVS1. However even if we turn-off the MW potential completely, the offset in the point of the Galactic plane crossing $X_{\mathrm{pl}},Y_{\mathrm{pl}}$ for orbits that backtrack from the current phase space position of the star to approximately the GC is a mere $\sim 15$\,pc, which is less than the width of the distribution $X_{\mathrm{pl}}$,$Y_{\mathrm{pl}}$. Similarly, setting the MW disk mass to zero causes a shift in $X_\mathrm{pl},Y_\mathrm{pl}$ of  $\sim 13$\,pc. If we take the potential of \citet{Mcmillan17} and vary the flattening (in density) of the Navarro-Frenk-White \citep{Navarro1997} DM halo from default $q_\mathrm{DM,halo}=1$ to $0.5$ or $2$ that results in offsets of only $\sim 7$ and $18$\,pc. 

This lack of sensitivity to the DM halo flattening was confirmed when we obtained the formal posterior on ${\mathcal P}(q_\mathrm{DM,halo}|D)$ under the hypothesis of ejection from the GC and found that it was consistent with the prior. This shows that with the current proper motion precision, S5-HVS1 cannot yet be used  to constrain the MW gravitational potential. One additional reason for the current lack of constraining power from S5-HVS1 is that, because we do not know the actual ejection velocity from the BH, we do not constrain the total deceleration of the star, but only deviations of the trajectory from a straight line. For meaningful potentials consistent with the existing data, the deviations from a straight line for a $\sim 2000$\,\kms star flying for $\sim 5$\,Myr are within a few tens of parsecs (listed above) and thus within the current uncertainties of the S5-HVS1 trajectory. With the improvement in proper motion precision from future \gaia data we expect, however, that constraints on the MW halo flattening will be possible.

\subsection{S5-HVS1 ejection by Sgr A*}

Given an almost certain GC origin of S5-HVS1, here we discuss possible implications for the ejection by Sgr A*. We focus on the \citet{Hills1988} mechanism involving a three-body interaction of a stellar binary with the SMBH leading to one star being ejected. There are other mechanisms involving binary black holes \citep{Yu2003,Levin2006} and a SMBH surrounded by a cluster of stellar mass black holes \citep{OLeary2008}, and we will discuss some of them later.

The first question we address is what are the expected properties of the binary required to produce the very high ejection speed of S5-HVS1.  To infer this we use the results of \citet{Bromley2006}, who parameterized the distribution of ejection velocities as a function of the black hole mass and binary parameters \citep[see equations 1-4 of][]{Bromley2006}. We adopt a black hole mass of $4.1\times10^6$\,M$_\odot$ from \citet{Gravity2018}, fix the mass of S5-HVS1 to the observed value $2.35$\,M$_\odot$ (see Table~\ref{tab:sed_fitres}) and adopt an ejection velocity of $1798\pm 3$\kms.  The remaining parameters required to compute the ejection velocity distribution are the semi-major axis of the binary $a$, the mass of the second star $M _2$ and the minimum approach distance $R_{\mathrm{min}}$ between the binary and the SMBH. We adopt a log-uniform distribution over the binary separation and the \citet{Chabrier2005} IMF prior on the mass of the secondary, and  $\pi(R_{\mathrm{min}})\sim R_{\mathrm{min}}$  prior for the minimum approach distance \citep[see][for details]{Bromley2006}. We require that the semi-major axis of the binary is larger than 2.5 $R_\odot$, which is approximately the expected radius of a star with a mass of $\sim 2.35$\,M$_\odot$ \citep{Boyajian2013}, and that the radius of S5-HVS1 is smaller than its tidal radius at the closest approach between the binary and the SMBH ($R_{\mathrm{min}}$). This limits the minimal separation of the binary and the SMBH $R_{\mathrm{min}}$ to be $\gtrsim 1.4$\,au.

\begin{figure}
    \centering
    \includegraphics{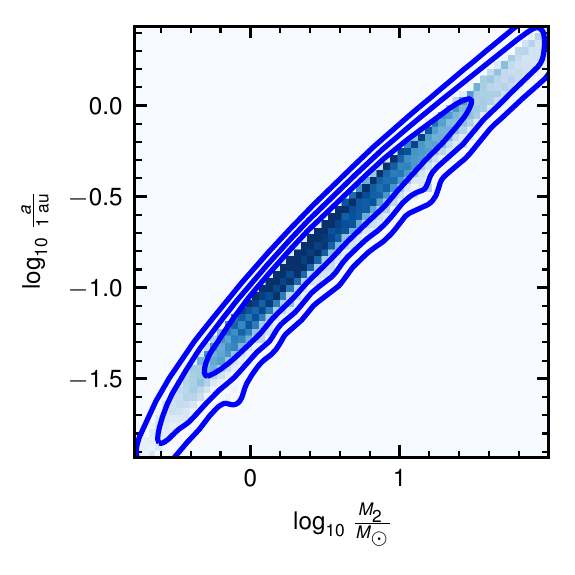}
    \caption{The distribution of semi-major axis of the binary system and mass of a secondary that could have produced S5-HVS1 via the Hills mechanism. The contours encircle the 68\%, 95\% and 99.7\% posterior volumes. }
    \label{fig:semimajor}
\end{figure}

Figure~\ref{fig:semimajor} shows our inferred probability distribution of the semi-major axis of the binary and mass of the second star. The distribution shows that in order to produce S5-HVS1  we need a former binary companion with mass $0.9$\, M$_\odot \lesssim M_2 \lesssim 16$\,M$_\odot$, where low  mass secondaries require an extremely tight separation of only $\sim 0.06$\, au, while if the secondary is massive, the semi-major can be as much as $\sim 0.63$\,au. The orbital periods of these binaries would range from 3 to 40 days. These ranges correspond to the 68\% confidence interval of the posterior. The binary parameters that we obtain are certainly possible \citep[see e.g.][]{Raghavan2010,Moe2013}, however, we expect these binaries to be quite rare.

\subsection{S5-HVS1 and stars around Sgr A*}

Given the certainty of the S5-HVS1 association with the Galactic Centre, it is interesting to assess if S5-HVS1 is related to any other structures known around the GC. 
The main stellar structure near the centre of the MW is the nuclear star cluster \citep{Becklin1968,Launhardt2002} with the Sgr A* SMBH at the centre. The central part of the star cluster consists of the so-called S-stars whose dynamics are dominated by the SMBH, and that orbit around it with periods from a few years to a few hundred years \citep{Ghez2005,GIllessen2009}. These stars are known to be massive and young \citep{Genzel2010,Lu2013} and we do not yet know how they came to be where they are. Furthermore, the cluster of stars around Sgr A* is known  to have substructure in the form of a coherently rotating small disk of young stars (the so-called clock-wise or CW disk) \citep{Levin2003,Paumard2006,Bartko2009,Yelda2014,Gillessen2017}.

The reason why S5-HVS1 can be potentially associated to some structures in the centre is that if the star has been produced by the Hills mechanism, then we expect that the direction of HVS's flight should be approximately aligned \citep{Lu2010,Zhang2010} with (i) the orbital plane of the original binary around the SMBH; and (ii) the orbital plane of the secondary star captured by the SMBH after the binary disruption (unless the secondary was swallowed by the black hole and/or produced a tidal disruption event). Thus we can hope to either identify a possible progenitor population of the S5-HVS1 binary or perhaps directly pinpoint the star that was previously paired to S5-HVS1 and still orbits Sgr A*. 

\begin{figure}
\includegraphics{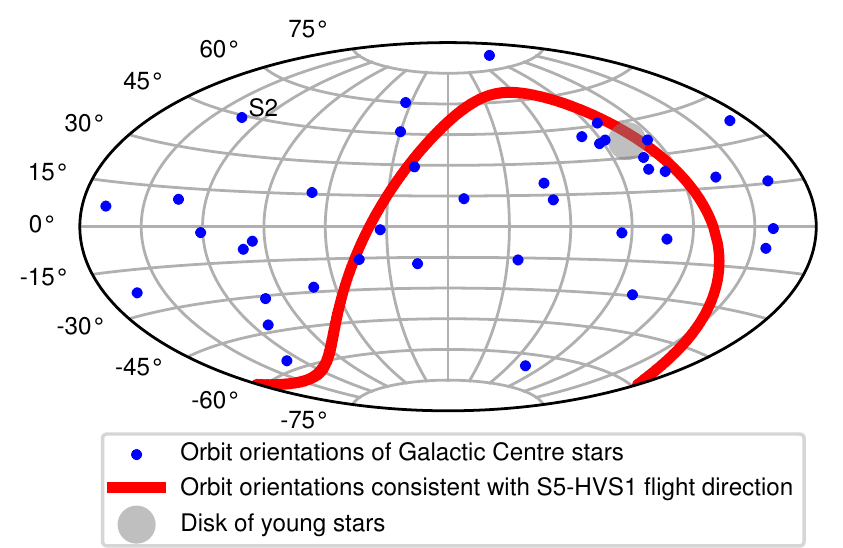}
\caption{
The orientation of orbital planes of stars around the GC compared to the possible orbital plane of the binary system around the GC that S5-HVS1 was member of. The coordinate system of the figure is the positional angle of ascending node of the orbit vs inclination of the orbit with respect to the line of sight. Blue circles identify orbits of stars from \citet{Gillessen2017}, while the red curve shows a set of possible planes consistent with S5-HVS1. The red curve also identifies the potential orbital plane of the secondary star of S5-HVS1 binary if it still orbits the SMBH and if S5-HVS1 was ejected by Hills mechanism. The grey circle marks the overdensity of stars on an orbital plane associated with the disk of young stars \citep{Bartko2009,Yelda2014}.} 
\label{fig:sgrstars}
\end{figure}

To check for possible association with the S-stars, we consider a set of possible orbital planes around the black hole that are aligned with the S5-HVS1 direction of flight. 
This set is clearly a 1-D manifold as there are infinitely many planes aligned with the vector pointing from the GC to S5-HVS1. 
On Figure~\ref{fig:sgrstars} we show the distribution of poles (or angular momentum directions) for this set of orbits by a red curve.  The coordinate system of the figure is the positional angle of the ascending node of the orbit and the angle between the orbital plane with respect to the vector from the Sun to the GC. Therefore the orbits seen edge-on from the Sun would occupy the equator on the figure, while face-on orbits would correspond to either the north or the south pole of the figure depending on the direction of rotation. On the figure we also overplot by blue circles the orientations of orbits (specifically the direction of their angular momenta) for stars around Sgr A* from \citet{Gillessen2017}. Thus if the S5-HVS1 has been produced by the Hills mechanism and the secondary star was captured on an orbit  around the BH, then the red curve should pass near the current orbital plane of the secondary star.  On the figure we also mark  by a grey circle the concentration of orbital poles corresponding to the disk of young stars observed around the GC \citep{Bartko2009,Yelda2014}. We see that the red line comes close to many S stars, which is not surprising and is expected to happen by chance. However, we see that the red curve also  crosses the concentration of blue points marked by the grey circle, meaning that S5-HVS1 flies within the orbital plane of young stars around the GC. This is potentially very interesting, because it may mean that the binary responsible for S5-HVS1 has the same origin as the young stellar disk. Several formation scenarios for it exist, that either involve the infall of a gas cloud on the GC with subsequent star formation \citep{Bonnell2008}, or star formation in the disk around the SMBH \citep{Nayakshin2007}. This disk could then potentially be the source of the S5-HVS1 binary \citep[i.e. due to orbit instabilities discovered by][]{Madigan2009} and S5-HVS1 would provide us with an opportunity of studying the stars in disk without all the complexities of observing through tens of magnitudes of extinction. In this scenario, the secondary of the S5-HVS1 could still be in the disk and thus could be potentially identified.

Alternatively, the young disk may consist of captured secondary stars from binaries disrupted in the Hills mechanism, in which case the previous partner of S5-HVS1 may be still in that disk. We note, however, that the young stellar age of the stars in the disk of a few Myr \citep{Lu2013}, and the low eccentricities of stars in the disk \citep{Yelda2014} make this scenario unlikely because the captured stars in Hills mechanism are expected to have high eccentricities \citep{Hills1988}.

\section{Discussion}
\label{sec:discussion}

\begin{figure}
\begin{center}
\includegraphics[]{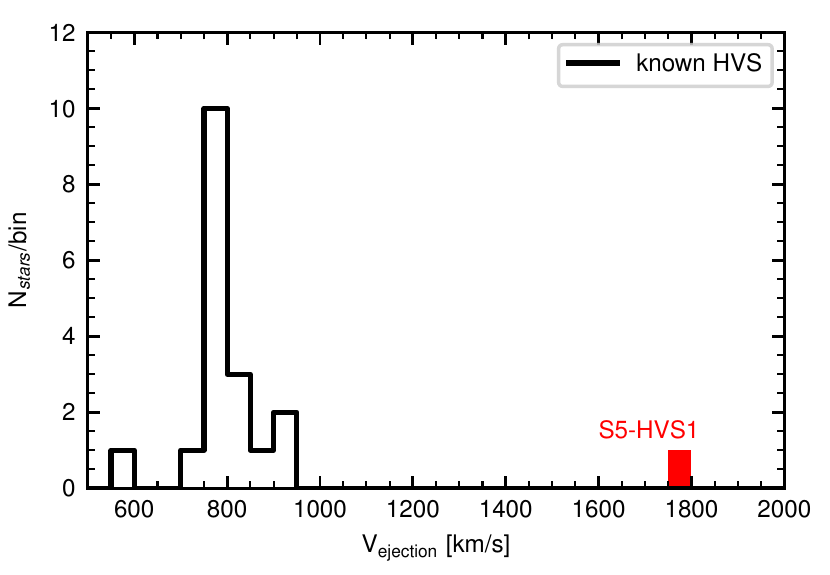}
\end{center}
\caption{The distribution of possible ejection velocities from the Galactic Centre computed for the subset of known HVS from \citet{Boubert2018} whose phase space measurements (position, velocities, proper motions and distance) are consistent with a GC ejection. S5-HVS1 is highlighted in red.}
\label{fig:vejection_distr}
\end{figure}

Here we address the multiple open questions that the discovery of S5-HVS1 poses. 
First, we compare S5-HVS1 to the other HVS. The main property that distinguishes S5-HVS1 from the rest of the hyper-velocity stars is its unusually high velocity. If we exclude the recently discovered D$^6$ white dwarfs produced in SNIa-like explosions \citep{Shen2018}, the velocity of S5-HVS1 is almost a factor of two larger than the velocity of any other known HVS. Figure~\ref{fig:vejection_distr} shows the distribution of likely ejection velocities from the Galactic Centre for other HVS. Here we use the same set of stars from \citet{Boubert2018} as shown on Figure~\ref{fig:finetuning}, and select a subset of those which can be well described ($\chi^2<20$) as being ejected from the Galactic Centre based on proper motion, position, distance and radial velocity. The figure shows how much of an outlier S5-HVS1 is, in particular because of the apparent clumping of previously known HVS at 800--1000\,\kms, which begs the question whether S5-HVS1 was produced using the same mechanism as other HVS. Another difference between S5-HVS1 and other HVS is that it is an A-type star, and thus is somewhat cooler, lower mass and later spectral type than the classical B-type hyper-velocity stars \citep{Brown2015}. It is also brighter and much more nearby than the majority of the faint, blue HVS that have been discovered in the Northern sky. 

One possible interpretation of these differences between S5-HVS1 and previously known HVS is that \SSSSS was just very lucky to stumble on a very rare object.  However the other explanation may be related to the somewhat lower mass and redder colour of S5-HVS1 $g-r=-0.27$, which is close to the colour boundary $g-r\sim-0.3$ of dedicated searches \citep{Brown2006,Brown2009}; this boundary minimises contamination because MS and BHB stars start to overlap at this colour. This may be the reason why previous spectroscopic searches missed lower mass/redder stars like S5-HVS1. However, since the Sloan Digital Sky Survey \citep[SDSS;][]{York2000,Yanny2009} did observe a large number of blue A-type stars in the range $-0.4<g-r<0$ and did not find anything close to S5-HVS1, it is useful to compare the number of objects spectroscopically observed by SDSS to \SSSSS. In fact, surprisingly, SDSS \citep[DR9;][]{SDSSDR9} has observed spectroscopically only $\sim$ 7 times more blue, distant (with small parallax $\pi<3\sigma_\pi$) stars in the $-0.4<g-r<-0.2$ and $16<g<18$  colour-magnitude range than \SSSSS did (1445 vs 202). Thus the SDSS non-detection of an S5-HVS1-like star is not in significant disagreement with the \SSSSS discovery.

Another possible explanation of the S5-HVS1 discovery has to do with its proximity, as the star is closer by a factor of several compared to other HVS. Why would closer HVSs be potentially noticeably faster or have a different velocity distribution? For this to happen it would require that the ejection mechanism of HVS is not operating at a constant rate and/or doesn't eject the same spectrum of HVS over time.  In the canonical \citet{Hills1988} mechanism where the loss cone of the SMBH is populated by slow scattering processes, such rapid changes would be problematic. However, as the presence of young (only few Myr old) stars and substructures near the GC indicate, the Galactic Centre has had a very active recent history; e.g.\ it is likely that the GC had an accretion event of a giant molecular cloud a few Myr ago that formed new stars \citep{Bonnell2008,Lucas2013} that were then distributed in a disk around the SMBH. If that is the case, that accretion event could have been a source of binaries for the Hills mechanism, producing an excess of stars in the orbital plane of accretion and an increased rate of HVS ejections a few Myr ago. In such a scenario, HVSs like S5-HVS1 could serve as timers and indicators of orientation of large accretion events happening near the GC. To test this hypothesis we will, however, need to find more stars with similar travel times as S5-HVS1. It is remarkable that the age of S5-HVS1 ejection is close to both the age of the disk of young stars around the GC \citep{Lu2013} and the age of the Fermi bubbles \citep{Bland-Hawthorn2003,Su2010} which have been potentially associated with the recent accretion event in the Galactic Centre \citep{Zubovas2011,Guo2012}, thus potentially linking these different astrophysical objects. 

An alternative scenario that would naturally produce a time-variable HVS spectrum is that involving an Intermediate Mass Black Hole (IMBH) orbiting the GC \citep{Yu2003,Levin2006}. In this mechanism, during the inspiral of the IMBH, the HVS production rate peaks and then subsides due to dynamical friction around the SMBH \citep{Baumgardt2006,Darbha2019}, with the fastest HVS being ejected in the final phase of the in-spiral. This mechanism produces a strongly anisotropic distribution with the fastest stars in the orbital plane of the IMBH \citep{Rasskazov2019}. There is also some indication that the HVS produced by this mechanism tend to have higher velocities and a flatter velocity spectrum than the classical  Hills mechanism \citep{Sesana2007}.
While there is currently not much evidence for the presence of an IMBH in the GC \citep{Gualandris2009} other than a shallow stellar density slope that can be produced by an IMBH scattering \citep{Baumgardt2006}, if an IMBH inspiral happened a few Myr ago, then it would produce an excess of nearby and fast HVSs with a narrow range of ejection times. To test for this possibility we need to search for other nearby HVS and see if there is an excess of stars that were ejected at roughly the same time as S5-HVS1 ($\sim 5$ Myr ago), are strongly anisotropic and that have a velocity spectrum inconsistent with the Hills mechanism.

An interesting consequence of the fact that S5-HVS1 has lower mass than most other HVS in the halo is that its expected life-time given the stellar mass of $\sim 2.3$\, M$_\odot $ is quite long -- around a $1$~Gyr. By the end of its life the star would have travelled a distance of $\sim $ 2\, Mpc,  traversing a large fraction of the Local Group. This suggests that searching for such  ejected stars at large distances from the MW or Andromeda \citep{Sherwin2008} is quite promising. On top of being well separated in colour-magnitude space from other contaminants, an S5-HVS1-like star would eventually evolve onto the red giant branch and thus be detectable much more easily. Searches for S5-HVS1-like stars within the whole Local Group will be possible with upcoming deep imaging surveys like LSST \citep{LSST2009}.

In this paper we tried to use the position-velocity information on S5-HVS1 to constrain the distance from the Sun to the Galactic Centre and the Galactic Solar velocity. We have not been able to constrain those simultaneously, mainly due to the precision of the distance determination to S5-HVS1. However, in the future, the combination of such constraints from multiple S5-HVS1-like stars (see Figure~\ref{fig:geomtry_constraints}) will resolve the existing degeneracies and should provide extremely precise measurements of the geometric and kinematic Galactic parameters. We believe that with the upcoming \gaia\ DR3 as well as future spectroscopic surveys like WEAVE \citep{Dalton2014}, 4MOST \citep{dejong2014} and DESI \citep{DESI2016}, the discovery of more HVS similar to S5-HVS1 is guaranteed. Furthermore, while with S5-HVS1 we were currently not able to put constraints on the gravitational potential due to the very short flight time and loose proper motion constraint, with the next \gaia data release that will increase the proper motion precision by a factor of few as well as deliver new HVSs, we think we will be able to start constraining the potential with individual HVS as predicted by \citet{Gnedin2005}. 

One other interesting prospect for the future of HVS science that we did not explore in this paper, but which may be promising, is that HVS could become probes of substructure and particularly DM substructure in the Galaxy, similar to stellar streams \citep{yoon2011,Erkal2016} or lensing \citep{Vegetti2012}. The reason for this is that for HVS that were ejected from the GC we know the orbit exactly, as it must connect to the Galactic Centre. Thus if we imagine a large collection of HVS travelling throughout the Galaxy, we expect that some of those trajectories will be affected by various external perturbations, including massive perturbers such as the Large Magellanic Cloud or Sagittarius Dwarf Spheroidal Galaxy, but also potentially smaller DM halos and globular clusters in the halo. Although we expect the effect of these perturbations to be quite small due the high velocity of the stars, if we have enough of these stars and they have high accuracy phase-space measurements, then we could say something about the mass substructure in the Galaxy. As an example, a $10^8$\,M$_\odot$ point-mass perturbing a hyper-velocity star travelling at 2000\,\kms with an impact parameter of 0.5 kpc will produce a velocity offset of $\sim 1$\kms \citep{Binney2011} perpendicular to the trajectory of the HVS, or equivalently an offset of $\sim $ a few parsecs in the trajectory. While these offsets are small, the velocity accuracy is within what \gaia proper motions will provide  for objects brighter than $G\sim 17$ within 10\, kpc.

Finally, let us consider the effect of future \gaia data releases on S5-HVS1. The main improvement will come from much higher precision parallax and proper motions, which are expected to better constrain the orbit of S5-HVS1. In advance of \gaia\ DR3, we predict that the true proper motions and parallax of S5-HVS1 are $\mu_\alpha \cos \delta=35.333\pm 0.080$\masyr, $\mu_\delta =0.617\pm 0.011$\masyr and $\varpi=0.11$\,mas (corresponding to a distance of $8.828$\,kpc). Time will tell whether these predictions based on the assumption of a GC origin will hold.

\section{Conclusions}
\label{sec:conclusions}

\begin{itemize}
    \item Using data from the \SSSSS spectroscopic survey we have identified a star with a radial velocity of $\sim 1020$ \kms without any signs of binarity across a year of observations.
    \item Analysis of the spectra and photometry of the star shows that it is likely an A-type $\sim 2.35$\,M$_\odot$ Main Sequence metal-rich star at a distance of $\sim$ 9 kpc. 
    \item Given the measured distance, proper motion and radial velocity, the total velocity of the star in the Galactic rest frame is $1755_{-45}^{+55}$\kms, making it the third fastest hyper-velocity (unbound) star in the Galaxy after the D$^6$ white dwarfs \citep{Shen2018}. 
    \item Backtracking the current phase-space position of S5-HVS1 to the MW disk points at a small elongated region of $\sim 50 \times 1000$\,pc$^2$ that contains the Galactic Centre. This provides incredibly strong evidence that the star was ejected from the Galactic Centre at speed of $\sim 1800$\,\kms around $\sim$ $4.8$\, Myr ago.
    \item If S5-HVS1 was ejected from the GC then we can constrain the distance to the Galactic Centre and the Solar velocity. If we assume the \citet{Gravity2019} prior on $R_0$, then our constraint on the y-component of solar velocity is $V_\odot=246.1\pm5.3\kms$, and, vice-versa, if the \citet{Gravity2019} prior is used on $V_\odot$, it leads to an $R_0$ constraint of $8.12\pm0.23$\, kpc. Due to the short flight time and non-negligible proper motion uncertainties, the star currently can not yet constrain the MW gravitational potential. 
    \item The direction of the S5-HVS1 ejection is curiously aligned with the disk of young stars around the Sgr A* suggesting a possible connection. This may mean that the star has been ejected in the same event that lead to the disk's formation. 
    \item The fact that S5-HVS1 was ejected with a velocity almost twice that of all other known HVS potentially originating from the GC poses two questions: were all the known HVS produced by the same mechanism and has the HVS velocity spectrum been constant in time?
\end{itemize}

\vspace{-0.2in} % to avoid empty page in the end
\section*{Acknowledgements}

This paper includes data gathered with Anglo-Australian Telescope in Australia under programme A/2018B/09. We acknowledge the traditional owners of the land on which the AAT stands, the Gamilaraay people, and pay our respects to elders past and present.

This work has made use of data from the European Space Agency (ESA) mission
{\it Gaia} (\url{https://www.cosmos.esa.int/gaia}), processed by the {\it Gaia}
Data Processing and Analysis Consortium (DPAC,
\url{https://www.cosmos.esa.int/web/gaia/dpac/consortium}). Funding for the DPAC
has been provided by national institutions, in particular the institutions
participating in the {\it Gaia} Multilateral Agreement.

% add grants here...
SK is partially supported by NSF grants AST-1813881, AST-1909584 and Heising-Simons foundation grant 2018-1030. 
DB is grateful to Magdalen College for his Fellowship by Examination and the Rudolf Peierls Centre for Theoretical Physics for providing office space and travel funds.
TSL and APJ are supported by NASA through Hubble Fellowship grant HST-HF2-51439.001 and HST-HF2-51393.001 respectively, awarded by the Space Telescope Science Institute, which is operated by the Association of Universities for Research in Astronomy, Inc., for NASA, under contract NAS5-26555.
JDS, SLM and DBZ acknowledge the support of the Australian Research Council (ARC) through Discovery Project grant DP180101791. G.~S.~Da~C. also acknowledges ARC support through Discovery Project grant DP1501013294. Parts of this research were supported by the Australian Research Council Centre of Excellence for All Sky Astrophysics in 3 Dimensions (ASTRO 3D), through project number CE170100013. We also thank the referee Ulrich Heber for a detailed report.

Software:
{\code{numpy} \citep{numpy}, \code{scipy} \citep{scipy}, \code{matplotlib} \citep{matplotlib}, \code{astropy} \citep{astropy:2013,astropy:2018}, \code{emcee} \citep{Foreman_Mackey2013}, \code{gala} \citep{gala}, \code{q3c} \citep{Koposov2006}, \code{isochrones} \citep{Morton2015}
\code{fastKDE} \citep{Obrien16}, \code{Dynesty} \citep{Speagle2019}, \code{pyMultiNest} \citep{Buchner2014}, \code{IPython} \citep{Perez2007}, \code{chainconsumer} \citep{Hinton2016}, \code{REBOUND} \citep{Rein2012},  \code{MultiNest} \citep{Feroz2009}, \code{RVSpecFit} \citep{rvspecfit19},
\code{schwimmbad} \citep{schwimmbad}}

\vspace{-0.2in} % avoiding empty space in the end
%%%%%%%%%%%%%%%%%%%%%%%%%%%%%%%%%%%%%%%%%%%%%%%%%%



\bibliographystyle{mnras}
\bibliography{refs.bib} % if your bibtex file is called example.bib


%%%%%%%%%%%%%%%%%%%%%%%%%%%%%%%%%%%%%%%%%%%%%%%%%%

%%%%%%%%%%%%%%%%% APPENDICES %%%%%%%%%%%%%%%%%%%%%

\appendix

% Don't change these lines
\bsp	% typesetting comment
\label{lastpage}
\end{document}